\documentclass[twocolumn]{aastex631}

\usepackage{amsmath}
\usepackage{bm}
\usepackage{diagbox}\usepackage{tabularx}
\usepackage{makecell}
\usepackage{xcolor}
\usepackage{nicefrac}

\graphicspath{{./}{Figures/}}

\begin{document}
\title{ {The Response of Planetary Atmospheres to the Impact of Icy Comets} I: Tidally-Locked {exo-Earths}}

\author{F. Sainsbury-Martinez}
\affiliation{School of Physics and Astronomy, University of Leeds, Leeds LS2 9JT, UK}
\author{C. Walsh}
\affiliation{School of Physics and Astronomy, University of Leeds, Leeds LS2 9JT, UK}
\author{G. Cooke}
\affiliation{Institute of Astronomy, University of Cambridge, UK}
\affiliation{School of Physics and Astronomy, University of Leeds, Leeds LS2 9JT, UK}

\begin{abstract} %250!
Impacts by rocky and icy bodies are thought to have played a key role in shaping the composition of solar system objects, including the Earth's habitability. Hence, it is likely that they play a similar role in exoplanetary systems. We investigate how an icy cometary impact affects the atmospheric chemistry, climate, and composition of an Earth-like, tidally-locked, terrestrial exoplanet, a prime target in the search for a habitable exoplanet beyond our solar system. We couple a cometary impact model which includes thermal ablation and pressure driven breakup with the 3D Earth System Model WACCM6/CESM2, and use this model to investigate the effects of the water and thermal energy delivery associated with an $R=2.5$ km pure water ice cometary impact on an Earth-like atmosphere.
We find that water is the primary driver of longer timescale changes to the atmospheric chemistry and composition by acting as a source of opacity, cloud ice, and atmospheric hydrogen/oxygen. The water opacity drives heating at $\sim5\times10^{-4}$ bar, and cooling below, due to a decreased flux reaching the surface. The increase in atmospheric hydrogen and oxygen also drives an increase in the abundance of hydrogen/oxygen rich molecules, with the exception of ozone, whose column density decreases by $\sim10\%$. These atmospheric changes are potentially observable for $\sim$ 1-2 years post-impact, particularly those associated with cloud ice scattering. They also persist, albeit at a much reduced level, to our quasi-steady-state, suggesting that sustained bombardment or multiple large impacts have the potential to shape the composition and habitability of terrestrial exoplanets.
\end{abstract}


\section{Introduction} \label{sec:introduction}

Material delivery associated with icy and rocky body impacts is proposed to have played a significant role in shaping the composition, and the habitability, of solar system planets. For example, material delivery by cometary and asteroidal impacts has been invoked in an effort to explain: i) Jupiter's super-solar metallicity, both for models which consider a more general enrichment \citep{2004jpsm.book...35G,2010SSRv..152..423F} and for models in which the outer atmosphere alone is enriched \citep{2024ApJ...967....7M}, ii) the dawn-dusk asymmetry of Mercury's exosphere \citep{BENZ1988516,Pokorny_2017}, and iii) the atmospheres of Mars \citep{WOO201987} and Venus \citep{2018SoSyR..52..392M,2023PhyU...66....2M}.\\
Cometary and asteroidal impacts have also been invoked as a key mechanism by which the early Earth's composition and atmosphere was shaped, potentially playing a critical role in the delivery of materials, such as complex organic molecules \citep{2002ESASP.518....9E} or water, which are key to setting the Earth's habitability \citep{DELSEMME2000313,2020AsBio..20.1121O}. For example, analysis of numerical models of the migration of small bodies from the outer solar system inwards towards the terrestrial planets suggests that the total mass of water delivered from the feeding zone of the giant planets ($\sim2\times10^{-4}\,M_\Earth$) to Earth is on the order of the mass of the Earth's oceans \citep{2003EM&P...92...89I,2004AdSpR..33.1524I,2014Icar..239...74O,2023PhyU...66....2M}, with similar relative water mass fractions also being delivered to Venus and Mars. Even the migration of dust from collisions of minor bodies throughout the solar system could play a significant role in delivering volatilities and organic molecules to the terrestrial planets \citep{Marov2005}.\\

Under the assumption that exoplanetary systems form in a similar manner to our own solar system, we can infer that planetary bombardment and cometary/asteroidal impacts should also have played a significant role in shaping the composition, atmosphere, and hence habitability of exoplanets. {For example, \citealt{2023RSPSA.47930434A}, explore the ability of cometary impacts to deliver prebiotic molecules to rocky/terrestrial exoplanets, suggesting that prebiotic molecules are more likely to survive impacts with low-mass planets orbiting high-mass stars due to the lower impact velocity of such a system. In a similar vein, \citet{2020A&A...638A..50F}, suggest that volatile delivery from exoplanetary system belts may have played a significant role in delivering volatiles to terrestrial planets that formed within the water snowline (i.e within a region of the protoplanetary disc that was too hot for water to condense and form ices), potentially delivering enough material to account for an Earth-like atmospheric mass. \citet{2022ApJ...937L..41C} confirm that such an impact driving belt might exist around an M-dwarf star, albeit with a lower occurrence rate than in higher-mass systems. Finally, \citet{2024ApJ...966...39S} discussed the role that icy cometary impacts might have in setting the observed low C/O ratios and high metallicities of hot gas giants. }  \\
This is particularly true for terrestrial planets orbiting cooler stars, such as low-mass M-dwarfs. The lower luminosities of these host stars result in the habitable zone - which is generally defined as the region around a star in which the equilibrium temperature of a planet would fall into the range that allows for liquid surface water without inducing a runaway greenhouse effect \citealt{2013ApJ...765..131K,2014A&A...567A.133V} - lying significantly closer to the host star than it would for a Sun-like star. An example is the TRAPPIST-1 system, where the habitable zone around this cool M8 red-dwarf host star lies between $\sim0.025$ and $\sim0.05$ au, which corresponds to to an orbital period of between $\sim4.5$ and $\sim13.5$ days \citep{Gillon2013,2016Natur.533..221G}. In turn, this leads to an increase in the orbital velocity of habitable-zone planets: the orbital velocity of TRAPPIST-1e is $\sim1.6$ times that of the Earth. This increase in orbital velocity, when combined with the effects of gravitational focusing by the nearby host-star \citep{2023PSJ.....4..139N}, suggests that impact rates for habitable zone planets orbiting M-dwarfs have the potential to be significantly higher than that experienced by the Earth, although this may be somewhat tempered by the lower protoplanetary disc masses associated with low-mass stars \citep{2017A&A...598L...5A}. {Note that, the multi-planet nature of the TRAPPIST-1 system may also lead to an enhanced impact rate. For example, \citet{2018MNRAS.473..295S} discuss the role that multi-planet resonances can have in increasing the planetesimal flux. } \\

In addition to its possible effect on the cometary impact rate, orbiting so close to a potentially active M-dwarf also may also have significant implications for the atmospheric chemistry and dynamics, effects which are likely to influence the response of a terrestrial planet's atmosphere to an icy cometary impact. \\
To start, the small orbital distance between such habitable-zone terrestrial planets and their hosts stars will lead to significant angular momentum exchange, via tidal torques, between the two bodies.
This will result in the synchronisation of the planetary rotation rate and orbital period, leaving us with a tidally-locked planet with a permanently illuminated day-side and permanently dark, and hence cooler, night-side \citep{1964hpfm.book.....D,2017CeMDA.129..509B}. In turn, this day-night insolation contrast leads to the formation of strong horizontal pressure and temperature gradients which, in concert with the somewhat rapid planetary rotation, shape and drive the global atmospheric circulations. 
For example, the strong day-night pressure/temperature gradient can lead to the formation of a global overturning circulation in which air parcels rise on the day-side, are advected towards the night-side in the stratosphere and above, and then sink surface-ward with near-surface night-to-day winds completing the circulation \citep{2013cctp.book..277S}. Simultaneously, the strong off-equator Coriolis effect can lead to the formation of standing Rossby and Kelvin waves which pump eastwards angular momentum from high latitudes to low, potentially driving to the formation of (a) super-rotating jet(s) \citep{2011ApJ...738...71S}.
Together, these circulations have the potential to significantly alter the overall climate of the planet \citep{2015MNRAS.453.2412C}, including the atmospheric chemistry and composition. A good example of this is atmospheric ozone, the formation and destruction of which is sensitive to UV insolation rates. Consequently, multiple studies have shown that day-night advection can lead to the enrichment of atmospheric ozone on the night-side and/or at the poles \citep[e.g.][]{2016EP&S...68...96P,Chen_2018,2020MNRAS.492.1691Y,10.1093/mnras/stad2704,cooke2024,anand2024}. \\
In addition to its implications for the orbital dynamics and atmospheric circulations, orbiting a cooler M-dwarf can also affect the atmospheric chemistry more directly due to differences in the stellar spectrum. Specifically, differences in the UV spectrum can drive significant changes in the atmospheric chemistry due to the sensitivity of many photochemical reactions, such as the formation of ozone or the photodissociation of water, to both the strength and shape (wavelength) of the incoming UV irradiation \citep[see, for example,][]{2014P&SS...98...66G,2018AsBio..18..630M,2022A&A...665A.156K}. \\

The importance of these effects emphasise the need to study the affects of cometary impacts on atmospheric dynamics and chemistry in a three-dimensional and time-dependent manner. {Here we conduct a pilot study, investigating a well documented parameter regime (a tidally-locked exo-Earth atmosphere) with a robust atmospheric model, laying the foundations for future studies which consider atmospheres more reminiscent of young terrestrial planets. Impact rates are expected to be much higher for young planets, such as the Archean-Earth \citep[e.g.][]{063016-020131,BRASSER2020113514}, and they likely played a significant role in delivering volatiles, such as oxygen or water \citep[e.g.][]{doi:10.1126/science.11538074,doi:10.1089/ast.2019.2187,2022PSJ.....3..115I}.  \\

To study such the effects of a single, icy, cometary impact,} we couple the cometary ablation and breakup model of \citet{2024ApJ...966...39S}, with WACCM6/CESM2, an Earth-System Model which has been used to explore the atmospheric dynamics and chemistry of both Earth-analogue \citep{10.1093/mnras/stac2604,2023MNRAS.524.1491L} and tidally-locked exoplanets \citep{2023ApJ...959...45C,cooke2024}.
{Then,} as discussed in \autoref{sec:method}, we use this coupled cometary-impact/climate model to study how the impact of a single pure-water-ice comet affects the atmosphere of a tidally-locked Earth-analogue exoplanet. Specifically, we consider an impact with TRAPPIST-1e which is the TRAPPIST planet with the highest likelihood of hosting a terrestrial, and hence potentially habitable, atmosphere \citep{2017ApJ...839L...1W}. \\
{An icy cometary impact affects the atmosphere of a tidally-locked exoplanet in two ways. Firstly it acts as a source of mass/water delivery, and secondly it delivers thermal energy to the atmosphere as the kinetic energy of the impacting comet is reduced. We discuss the combined effects of these two components of the cometary impact on our tidally-locked, Earth-like atmosphere in \autoref{sec:results_combined}. We also ran two additional models in which we isolate the two components of the cometary impact driven delivery. A discussion of these models, and the differences in both the strength and the timescale of the atmospheric response to isolated mass/water deposition (\autoref{sec:water_isolated}) and heat deposition (\autoref{sec:heat_isolated}), can be found in \autoref{sec:results_isolated}.  }
We finish, in \autoref{sec:concluding}, with some concluding remarks, discussing the implications of our results for our understanding of the compositions of terrestrial exoplanetary atmospheres as well as possible directions that this work {will} take in the future.

\section{Method} \label{sec:method}

To understand how an icy cometary impact affects the atmosphere of a terrestrial, tidally-locked, exoplanet, we couple a slightly modified version of the parametrised cometary impact model of \citet{2024ApJ...966...39S} (\autoref{sec:cometary_impact_model}) with a version of the Earth-System Model WACCM6/CESM2, which has been modified to allow for synchronous rotation\footnote{\url{github.com/exo-cesm/}} and the delivery of both thermal energy and water due to a cometary impact\footnote{\href{gitlab.com/leeds_work/cesm_comet}{gitlab.com/leeds\_work/cesm\_comet}} (\autoref{sec:atmosphere_model}). This coupled model is then used to study the impact of a pure water ice comet, with a radius of $2.5$ km and a density of $1\,\mathrm{g\, cm^{-3}}$, with the atmosphere of a quasi-steady-state TRAPPIST-1e model with pre-industrial atmospheric composition and an Earth-like land-ocean distribution, including orography (mountains) and a dynamic ocean. {We use this configuration becasue it has been robustly tested and benchmarked \citep{10.1093/mnras/stac2604,2023MNRAS.524.1491L,2023ApJ...959...45C,anand2024,cooke2024,sainsbury2024b}. } 

\subsection{Cometary Impact Model} \label{sec:cometary_impact_model}
\begin{figure}[tp] %
\begin{centering}
\includegraphics[width=0.98\columnwidth]{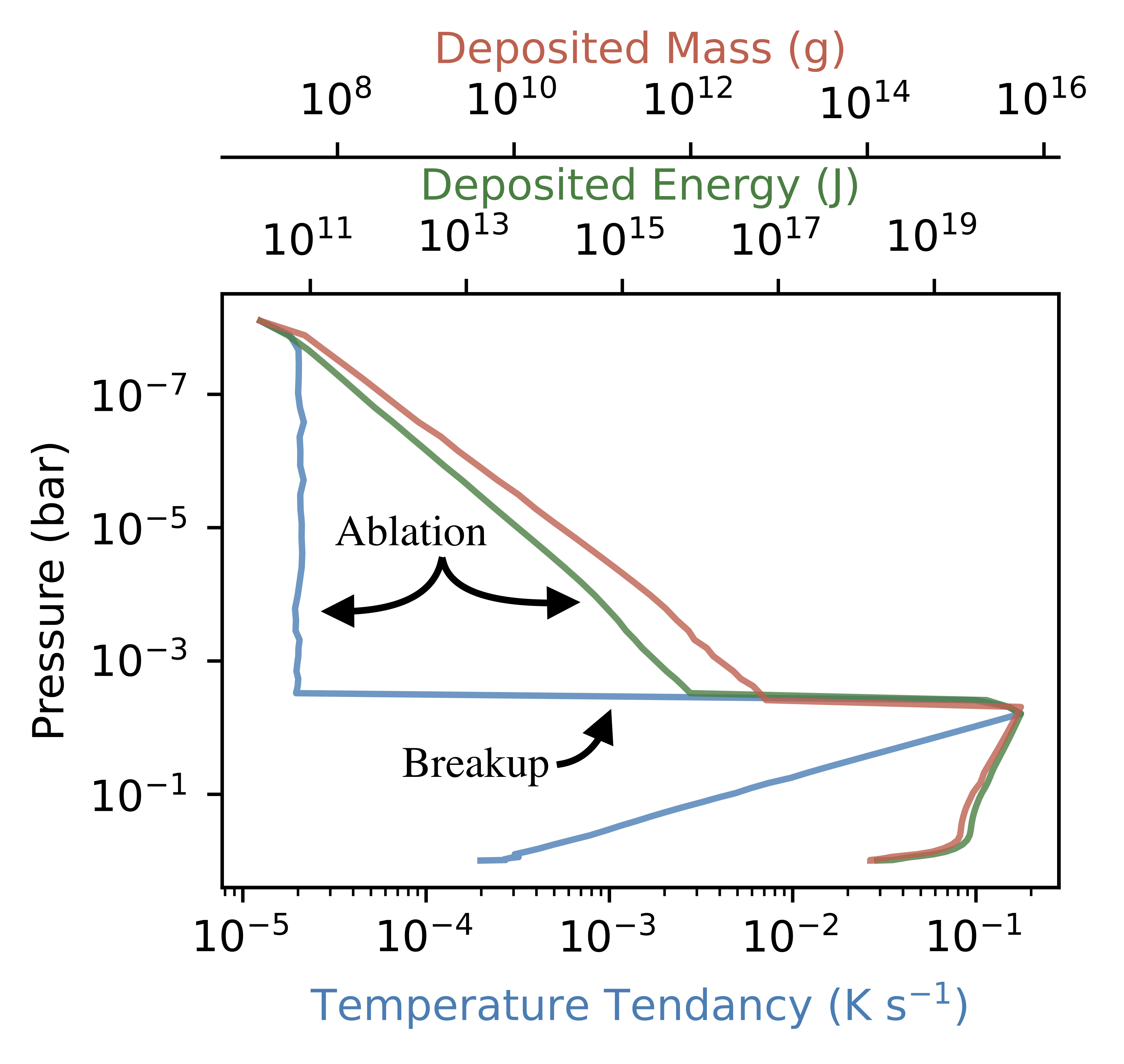}
\caption{ The initial vertical mass (red) and thermal energy distribution profile (green) generated by our cometary ablation and break-up model for a pure water ice comet, with a radius of 2.5 km and a density of $1\,\mathrm{g\, cm^{-3}}$, impacting the sub-stellar point of our TRAPPIST-1e-like atmospheric model. 
For computational reasons, this mass and energy is introduced to the model over a period of 10 days, and with an initial horizontal spread of nine (3x3) columns. Further, due to the structure of WACCM6/CESM2's solver, the deposited energy must be converted into a local temperature tendency (i.e. rate of temperature change) which is then applied to each cell (blue).  \label{fig:energy_mass_deposition} }
\end{centering}
\end{figure}

To model our icy cometary impacts, we adopt the parametrised cometary ablation and breakup model of \citet{2024ApJ...966...39S}, albeit with a modification to the thermal energy deposition during the ablation phase to address the lower heat capacity of the outer atmosphere of a terrestrial planet when compared with a hot gas giant. \\
This model assumes that the comet encounters the atmosphere with a zero angle of incidence (i.e. $\cos\left(\theta\right)=1$), that it remains spherical until breakup (i.e. no deformation), and that its interaction with the atmosphere can be split into two distinct phases: in the outer atmosphere, where the pressure/atmospheric-density is low, the comet slows due to atmospheric drag which drives surface {\it ablation} (phase 1). However as the density of the atmosphere increases, so too does the drag and stress on the comet, leading to an increase in the ram pressure and eventual cometary {\it breakup} (phase 2) when the ram pressure exceeds the tensile strength of the cometary ice \citep{PASSEY1980211,2016ApJ...832...41M}. Note that a number of other complex physical phenomenon are hypothesised to play a key role in cometary breakup, such as deformation (i.e. pancaking - \citealt{1994ApJ...434L..33M,1995ApJ...438..957F}) or surface instabilities (e.g. Rayleigh-Taylor - \citealt{2005ApJ...626L..57A} - or Kelvin-Helmholtz - \citealt{KORYCANSKY20021}), however, as discussed in \citet{2024ApJ...966...39S}, a careful consideration of ram driven breakup alone is sufficient to reproduce the breakup locations of observed cometary impacts, such as Shoemaker-Levy 9 on Jupiter. \\
During the ablation phase of the cometary impact we model the velocity ($V$) and location of the comet in a time-dependent manner via the velocity evolution equation of \citet{PASSEY1980211}:
\begin{equation}
    \frac{dV}{dt} = g-\frac{C_{D}\rho_{a}AV^{2}}{M}, \label{eq:velocity_evo}
\end{equation}
where 
\begin{equation}
    A=S_{F}\left(\frac{M}{\rho_{c}}\right)^{\frac{2}{3}}
\end{equation}
is the effective cross-sectional area of the spherical comet, $g=9.1454\,\mathrm{m\, s^{-2}}$ is the gravitational acceleration associated with the planet TRAPPIST-1e \citep{2021PSJ.....2....1A}, $C_{D}=0.5$ is the drag coefficient (of a sphere), $\rho_{a}$ is the atmospheric density ({taken from the input CESM atmospheric model}), $S_{F}=1.3$ is the shape factor (of a sphere), $dt$ is the timestep (which must be short enough to capture the impact), and $\rho$, $dV$, and $M$ are the density, change in velocity, and remaining mass of the comet respectively. 
Here we assume that the density of the comet is the same as pure water ice ($\rho=1\,\mathrm{g\, cm^{-3}}$), that the radius of the comet is 2.5 km (which falls on the lower end of known cometary radii - \citep{2011ARA&A..49..281A}), and that the initial velocity of the comet is $V=10\,\mathrm{km\, s^{-1}}$, the latter of which is approximately the escape velocity of TRAPPIST-1e. Note that we limit ourselves to smaller cometary impacts ($R=2.5$ km) due to computational instabilities that occur for higher mass/energy deposition rates. Hence the quantification performed here on the climate response to the impact can be considered a lower threshold since one would expect impacts from larger, more massive comets to have a greater influence on the atmosphere.  \\
This velocity is then used to calculate the ablation driven mass-deposition, which is given by
\begin{equation}
    \frac{dM}{dt} = -\frac{C_{H}\rho_{a}AV^{3}}{2Q}\left(\frac{V^{2}-V^{2}_{cr}}{V^{2}}\right), \label{eq:ablation}
\end{equation}
where $C_{H}=0.5$ is the heat transfer coefficient \citep{1995Icar..116..131S,2005A&A...434..343A}, $Q=2.5\times10^{10}\,\mathrm{erg\, g^{-1}}$ is the heat of ablation of the cometary ice (pure water; \citealt{2016ApJ...832...41M}), and $V_{cr}=3\,\mathrm{km\, s^{-1}}$ is the critical velocity below which no ablation occurs \citep{PASSEY1980211}\footnote{{ Note that this equation is incorrect in \citet{PASSEY1980211}, as well as a number of other works that have cited this work. Specifically, their version of \autoref{eq:ablation} (equation 2 in \citealt{PASSEY1980211}) is missing a factor of $V$ (i.e. $V^2$ vs $V^3$) in the first term.}}. \\
The thermal energy deposition in the outer atmosphere is given by
\begin{equation}
    \frac{dE}{dt} = 0.5\pi V^{3}\rho_{a}R^{2}. 
\end{equation}
Here, rather than directly depositing a fraction of the lost kinetic energy of the comet into the atmosphere we instead consider a reduced thermal input based upon interactions between the comet and the column of atmospheric gas it passes through. More specifically, we calculate the deposited thermal energy by working out how many `static' molecules interact with the comet each time step, with each molecule gaining kinetic energy based upon the instantaneous velocity of the comet,  i.e. by assuming that the comet scatters molecules out of the column it passes through with a velocity equal to the cometary velocity. Even then, as we discuss in \autoref{sec:atmosphere_model}, the amount of thermal energy added to the atmosphere is at limit of what our model can computationally handle.  \\
At the same time, we also calculate the ram pressure ($P_\mathrm{ram}$) and compare this against the tensile strength of the comet ($\sigma_{T}=4.6\times10^{6}\,\mathrm{erg\, cm^{-2}}$; the tensile strength of an icy planetesimal taken from \citealt{2016ApJ...832...41M}), with cometary breakup considered to have occurred when,
\begin{align}
    P_\mathrm{ram} &> \sigma_{T}, \mathrm{where}\\
    P_\mathrm{ram} &= C_{D}\rho_{A}V^{2}. 
\end{align}
Once this condition is fulfilled, the comet is destroyed and any remaining mass and kinetic energy is distributed deeper into the atmosphere over a pressure scale height. This is done using an exponentially decaying function which is normalised to ensure that all cometary material/deposited energy is introduced to the atmosphere between the break-up site and the surface. This simulates the rapid breakup of cometary ices and the resulting mass/energy distribution due to the inertia of the impacting material without the need for complex fragment modelling. \\
The resulting mass and thermal energy deposition profiles for the impact of a pure-water-ice comet with the sub-stellar point of TRAPPIST-1e are shown in \autoref{fig:energy_mass_deposition}. A full list of the cometary parameters used to calculate these ablation profiles are given in \autoref{tab:comet_parameters}. We discuss the background atmosphere with which these comets interact {(i.e. which is used to calculate the profiles and to which the profiles are applied)} below (\autoref{sec:atmosphere_model}).\\
Here we can see the two stages of our cometary impact. In the low density outer atmosphere we find that both the mass (ablation) and thermal energy deposition rates increase with pressure, and hence atmospheric density. However there comes a point, around { $4.8\times10^{-3}$ bar ($\sim34$ km above the surface)} in our model, that the stresses on the comet have become too much, leading to break-up and the deposition of the majority of the comets ice and kinetic energy into the deeper atmosphere.
\begin{table}
\centering
\begin{tabular}{lcc}
Parameter & Value & Unit \\ \hline
Radius $R$ & $2.5$ & km\\
Density $\rho_c$ & 1 & g cm$^{-3}$ \\
Initial Velocity $V$  & 10 & km s$^{-1}$  \\
Heat Transfer Coefficient $C_{H}$  & 0.5 & -\\
Drag Coefficient $C_{D}$ & 0.5 & - \\
Latent Heat of Ablation  $Q$ & $2.5\times10^{10}$ & erg g$^{-1}$ \\
Tensile Strength $\sigma_{T}$ & $4\times10^{6}$ & erg cm$^{-2}$
\end{tabular}
\caption{Parameters of the impacting icy comet considered in this work.}
\label{tab:comet_parameters}
\end{table}
\subsection{Planetary Atmosphere Model} \label{sec:atmosphere_model}
In order to model the response of a tidally-locked, terrestrial, exoplanetary atmosphere to an icy cometary impact {we couple the above ablation and breakup model to a } TRAPPIST-1e-like atmospheric model calculated with the Earth-System Model WACCM6/CESM2. Additionally we also simulate an unimpacted reference case with which to compare our results, { and two additional coupled models, in which we explore the isolated effects of water/mass and thermal-energy deposition, and which are used to investigate the strength and life-time of the changes associated with each component of the cometary impact}.  \\

\subsubsection{WACCM6/CESM2}
The Whole Atmosphere Community Climate Model (WACCM6) is a well documented \citep{https://doi.org/10.1029/2019JD030943}, high-top (the atmosphere extends to 140 km - $\sim10^{-8}$ bar - above the surface), configuration of the open-source Coupled Earth-System Model (CESM2). It includes a modern (i.e. current-day), Earth-like land-ocean distribution with orography, and numerous initial atmospheric compositions, ranging from current day to pre-industrialisation, the latter of which we consider here. 
Horizontally, the simulation has a resolution of $1.875^{\circ}$ by $2.5^{\circ}$, which corresponds to 96 cells latitudinally (north-south) and 144 cells zonally (east-west). Vertically the simulation domain is split into 70 pressure levels distributed in $\log\left(P\right)$ space such that the number of pressure levels increases near the dynamically active surface. It has been modified by \citet{2023ApJ...959...45C} to account for the effects of synchronous rotation (see that work for more details). \\
Coupling WACCM6/CESM2 with our cometary impact model also required that we add two new external forcing sources to the model: water and thermal energy. In both cases these external forcing terms take the form of a rate of material or thermal energy input as well as a time-frame over which to apply this input. 
In the case of the water deposition we spread the material out both temporally, introducing the material over ten Earth days, and spatially, spreading the material over nine columns centred on the sub-stellar point. 
This `slow' and spread-out deposition is necessary in order to limit the water abundance gradient near the impact site. Higher abundance gradients induce numerical instabilities in the radiative transfer solver, leading to model termination. \\

The thermal energy deposition is a little more complicated to implement since CESM's physics solver does not accept a direct thermal power input. Instead the deposited thermal energy must be converted to a temperature tendency 
\begin{equation}
    \frac{dT}{dt} = \frac{dE}{dt} \cdot \frac{1}{\rho_{a}\mathcal{V}c_{p}}, \label{eq:temperature_tendancy}
\end{equation}
where $\mathcal{V}$ is the volume of the cell in which the thermal energy has been deposited, and $c_{p}$ is the specific heat capacity at constant pressure. To prevent the formation of large, numerically unstable, temperature gradients near the impact site, we spread this thermal energy out in the same way as was done for the deposited water. 
The final temperature tendency profile that is used as our cometary heat deposition profile is shown in blue in \autoref{fig:energy_mass_deposition}. \\
A more technical discussion of how water and thermal energy deposition is implemented in WACCM6/CESM2 can be found online\footnote{\href{gitlab.com/leeds_work/cesm_comet}{gitlab.com/leeds\_work/cesm\_comet}}.  \\
%%%%%%
\begin{table}[htp]
\centering
\begin{tabular}{lcc}
Parameter & Value & Unit \\ \hline
Radius $R$ & 0.91 & R$_\earth$\\
Mass $M$  & 0.772 & M$_\earth$ \\
Semimajor Axis $a$ & 0.0292 & au \\
Orbital Period $P_{orb}$ & 6.099 & days\\
Obliquity $\epsilon$ & 0 & \, \\
Eccentricity $e$ & 0 & \, \\
Peak Insolation $I$ & 900 & W\,m$^{-2}$ \\
Surface Gravity $g$ & 9.1454 & m\,s$^{-2}$
\end{tabular}
\caption{Planetary parameters of TRAPPIST-1e, taken from \citet{10.1093/mnras/sty051}, \citet{2018A&A...613A..68G}, and \citet{2021PSJ.....2....1A}, with the mass and radius of the planet chosen to be consistent with those from the TRAPPIST-1 Habitable Atmosphere Intercomparison (THAI) program \citep{2021PSJ.....2..106F,2022PSJ.....3..211T,2022PSJ.....3..212S,2022PSJ.....3..213F}. }
\label{tab:TP1E_parameters}
\end{table}
%%%%%%%
\subsubsection{TRAPPIST-1e}
Here, as our initial unperturbed reference state, we consider a TRAPPIST-1e-like planet with the sub-stellar point fixed over the pacific ocean (see Figure 1 of \citealt{sainsbury2024b}). Note that our model was evolved for over 300 years before impact so as to ensure that any effects associated with the atmospheric dynamics settling into a tidally-locked circulation regime have dissipated. \\
Briefly, TRAPPIST-1e is a terrestrial planet which remains a significant object of interest in the search for a habitable, Earth-like, exoplanet. It is slightly smaller and less massive than the Earth, with a radius of $0.91\,\mathrm{R_{\earth}}$ and a mass of $0.772\,\mathrm{M_{\earth}}$, leading to a slightly weaker surface gravity of $9.1454$ m\,s$^{-2}$. It also orbits significantly closer to its host star than the Earth does the Sun, with an orbital period of only $\sim6.1$ days. However, because TRAPPIST-1 is a cool M-dwarf, the peak insolation that TRAPPIST-1e receives (900 W\,m$^{-2}$) is around 66\% of that received by the Earth, placing it near the cooler edge of the habitable zone. To reproduce this insolation in our models we follow the work of \citet{2023ApJ...959...45C} and rescale the TRAPPIST-1 spectrum of \citet{2019ApJ...871..235P} such that the total, integrated, insolation matches that of TRAPPIST-1e. A summary of the planetary parameters of TRAPPIST-1e is given in \autoref{tab:TP1E_parameters}.

\begin{figure*}[tbp]
\begin{centering}
\includegraphics[width=0.85\textwidth]{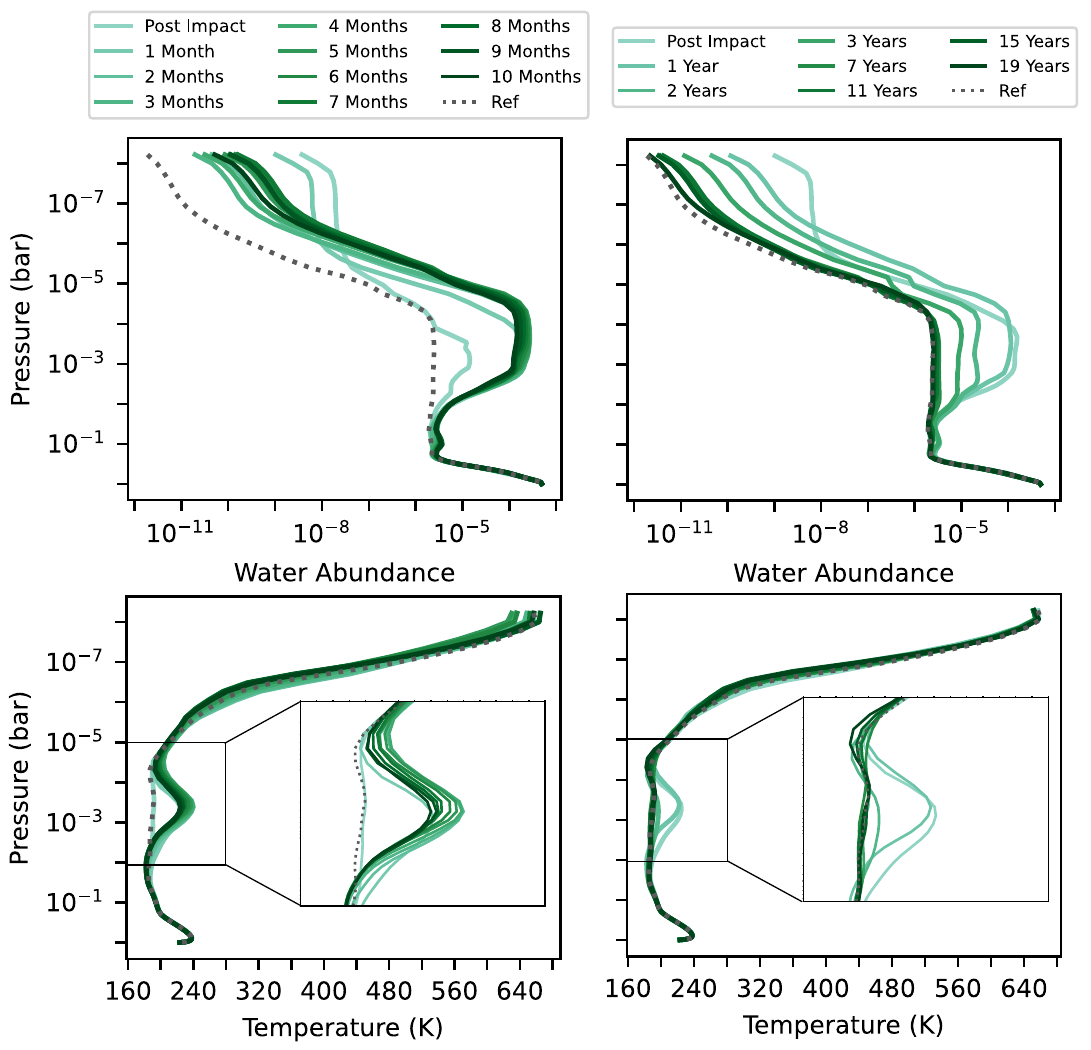}
\caption{ Fractional water abundance (top) and temperature (bottom) profiles showing the rapid atmospheric evolution within the first 10 months (left) of the impact of a pure water ice comet and the slower but steady settling of the atmosphere into a quasi-steady-state (right) reminiscent of the non-impacted reference state (grey dashed). Here each profile is calculated by averaging both horizontally over all latitudes and longitudes, and temporally over a month of simulation time. To better demonstrate the change in temperature in the mid-atmosphere, we include an inset showing a zoomed-in view of the temperature profile between $10^{-2}$ and $10^{-5}$ bar.   \label{fig:combined_curves_water_temp} }
\end{centering}
\end{figure*}
\begin{figure*}[tbp]
\begin{centering}
\includegraphics[width=0.99\textwidth]{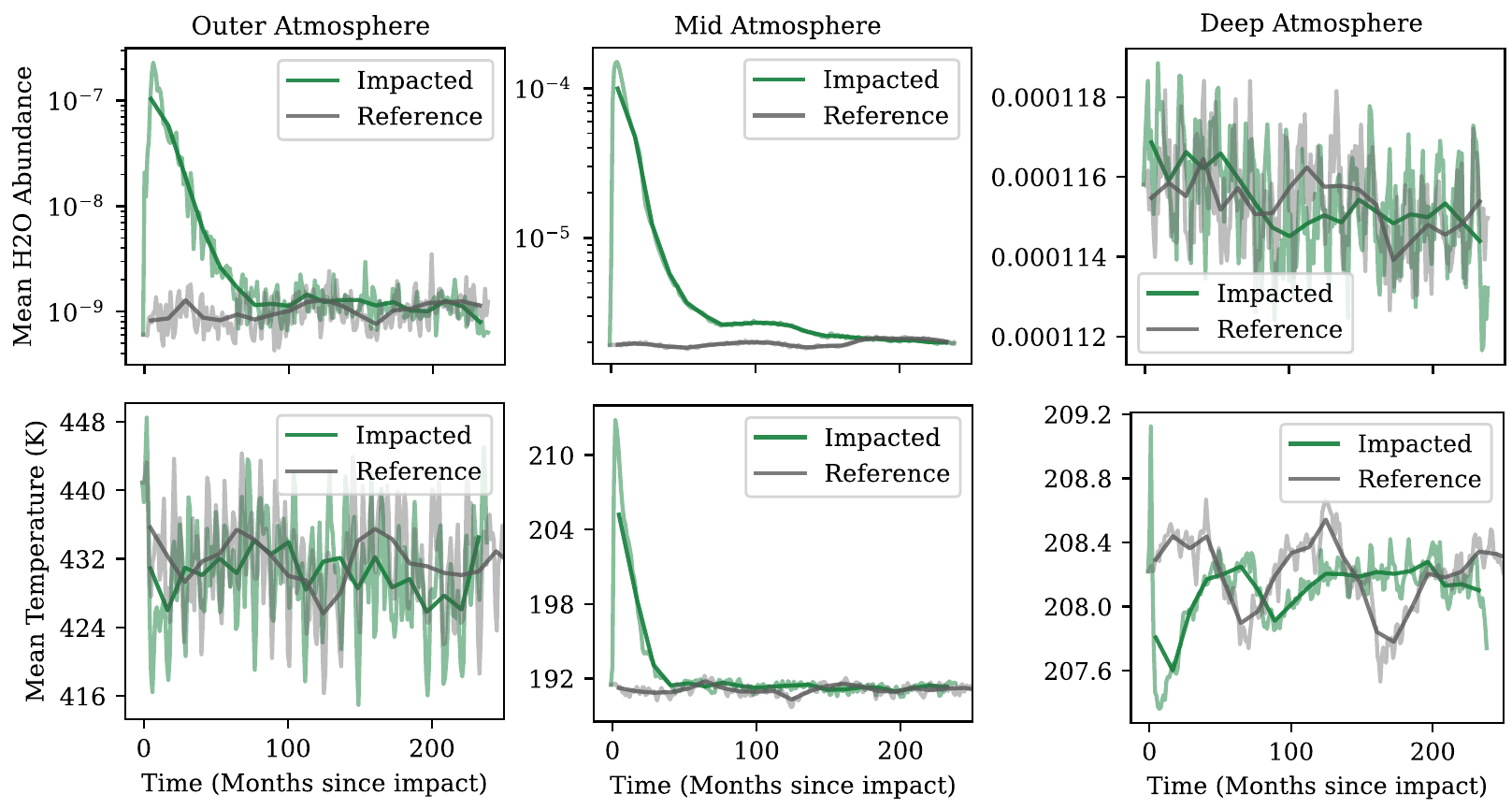}
\caption{ Time evolution of the annual mean (solid lines) and monthly mean (faint lines) fractional water abundance (top row) and temperature (bottom) in the outer atmosphere ($P<10^{-5}$ bar - left), mid-atmosphere ($10^{-5}>P>10^{-2}$ bar - middle), and near the surface ($P>10^{-2}$ bar - right) {of our fiducial coupled model (i.e. both water and thermal deposition - green)} and our non-impacted reference state (grey).    \label{fig:comparison_time_evo_curves_water_temp_combined} }
\end{centering}
\end{figure*}
\section{Results} \label{sec:results_combined}

{Our analysis of the isolated effects of cometary water and heat deposition (see \autoref{sec:results_isolated}) suggested that, unlike in our previous hot Jupiter studies \citep{2024ApJ...966...39S}, both components of the cometary material deposition can play a significant role in shaping the post-impact planetary atmosphere. As such, here we focus our discussion on a fiducial model which} couples both components of the icy cometary impact with our tidally-locked, terrestrial, exoplanetary atmosphere.

\subsection{Water Abundance and Mean Temperature} \label{sec:fiducial_evo}

{\autoref{fig:combined_curves_water_temp} shows how the fractional water abundance (top row) and mean temperature (bottom) vary both shortly after impact (left column) and over the $\sim20$ years required for the model atmosphere to reach a quasi-steady-state (right column). Note that the increase in fractional water abundance between the `Post-Impact' and `1 month' profiles in \autoref{fig:combined_curves_water_temp} occurs because we are plotting monthly mean values and the impact occurs part way through a month, leading to a reduced temporally averaged value. \\ 

Starting with the fractional water abundance, the top row of \autoref{fig:combined_curves_water_temp} shows how even a single icy cometary impact can change the atmospheric water content, particularly at low pressures, and how these changes can persist for years post impact. These profiles also reveal the role that vertical transport plays in the distribution of cometary material. As can be seen in \autoref{fig:combined_curves_water_temp}, most of the cometary water is delivered at pressures $>5\times10^{-3}$ bar, yet after only one month of simulation time, we find a { several order-of-magnitude increase in water abundance for all pressures $<10^{-3}$ bar. And we know that this enhancement is associated with vertical transport as the initial thermally ablated water shows signs of rapid settling at low pressures before mixing from the mid-atmosphere replenishes this reservoir (see the top left panel of \autoref{fig:combined_curves_water_temp} which reveals a decrease in outer atmosphere water abundance in the first few months post impact).} \\

On the other hand, the atmosphere near the surface exhibits almost zero mean response to the influx of water, despite the relatively weak drop-off in deposited mass between the break-up site and surface (\autoref{fig:energy_mass_deposition}). This can be attributed to the density of the atmosphere, which increases rapidly as we approach the surface. As such, the same mass of water has a much smaller effect on the atmospheric composition near the surface compared to low pressures. However the limited effect of cometary water deposition on the near-surface atmosphere does not mean that cometary impacts will not be observable as transmission spectra typically probe low-pressure regions, like the outer and mid-atmosphere where the effects of water deposition are largest (see \autoref{sec:observational}). \\

The difference in response to the cometary water deposition of different layers of the atmosphere can also be seen in \autoref{fig:comparison_time_evo_curves_water_temp_combined}. Here we plot the temporal evolution of the mean fractional water abundance and temperature in the outer atmosphere ($P<10^{-5}$ bar - left), the mid-atmosphere ($10^{-5}<P<10^{-2}$ bar - middle), and near the surface (referred to as the deep-atmosphere with $P>10^{-2}$ bar - right). These pressure regions were chosen in order to emphasise the pressure dependence of the atmospheric response to a cometary impact. By comparing these profiles we can see how the largest `spike' in fractional water abundance can be found in the outer atmosphere; over { two} orders of magnitude { three months} post impact, with significant enhancements persisting for at least 5 years post-impact. On the other hand, the longest lasting enhancement in atmospheric water can be found in the mid atmosphere, with the fractional water abundance remaining significantly enhanced with respect to that found in our non-impacted reference case for over 15 years post impact. Finally we also find a very slight enhancement in the near-surface water abundance shortly after the impact. Our analysis of the thermal energy deposition in isolation (\autoref{sec:heat_isolated}) suggests that { one} driver of this { deeper} enhancement is an initial burst of cometary heating driven evaporation/sublimation of water-droplets, ice, and snow, combined with the relatively weak water deposition. \\

The influx of a significant fraction of the cometary ice/water and thermal energy into the mid-atmosphere also has a significant effect on the local temperature. As shown on the bottom row of \autoref{fig:combined_curves_water_temp}, we find that a thermal inversion forms at $\sim 5\times10^{-4}$ bar (i.e. at the top of the stratosphere). 
At its peak in the resolved (i.e. non-averaged) data, which is located at the sub-stellar point, this temperature inversion is { 35} K hotter than the same location in our non-impacted reference case, and on average the mid-atmosphere is { $\sim 15$} K hotter (\autoref{fig:comparison_time_evo_curves_water_temp_combined}). This heating occurs due to the opacity of the deposited water \citep{2000ApJ...537..916S,2008ApJ...678.1419F,2010ApJ...719..341B,2024ApJ...966...39S} which causes an increased fraction of the incoming irradiation to be absorbed in the mid-atmosphere, driving localised heating. \\
This enhanced mid-atmosphere opacity, combined with albedo/scattering associated with clouds/ice (\autoref{sec:water_implications}), also affects the near-surface temperature (\autoref{fig:comparison_time_evo_curves_water_temp_combined}). We find that, despite the large amounts of thermal energy that the cometary break-up delivers (\autoref{fig:energy_mass_deposition}/\autoref{sec:heat_isolated}), the reduction in flux reaching the (near)-surface drives a $\sim1$ K decrease in temperature. This cooling might have implications for planetary habitability in a theoretical scenario in which cometary impacts are sufficiently regular that the atmosphere remains optically thicker or has a higher albedo.\\ %Note however that, whilst the near-surface changes associated with a single impact remain small, on the order of the `monthly' variability in the deep fractional water-abundance and mean temperature, the impact has disrupted the delicate multi-eyar cycles that operate within our tidally-locked exoplanetary atmosphere, leading to the out of phase oscillations seen in \autoref{fig:comparison_time_evo_curves_water_temp_combined}.  \\
As for how the thermal energy from the cometary impact has dissipated into the atmosphere, beyond a very short lived temperature spike { in the deep atmosphere} (less than one month), most of it has contributed towards either balancing the aforementioned drop in insolation or driving water evaporation/sublimation. The latter effect can be seen when comparing the mean fractional water abundance in our fiducial model with its isolated deposition counterparts (\autoref{sec:results_isolated}). \\
Post impact, we also find that the temperature of the outer atmosphere is slightly lower than our non-impacted reference model. This effect can be linked with the slight shift in the thermosphere to lower-pressures/higher-altitudes, an effect which itself is driven by the warming, and hence expansion, of the mid-atmosphere.  \\

Finally, we note that the changes to the global mean temperature of the atmosphere are much shorter lived than the changes to the fractional water abundance which drive them. 
This occurs due to a combination of the tidally-locked nature of the illumination and the advection of water from the impact site (\autoref{sec:advection}). 
The tidally-locked illumination means that water-opacity effects are strongest on the day-side, particularly at the sub-stellar point. Whilst advective mixing of water vapour leads to the transport of deposited water from the insolated day-side to the dark night-side where it has only a very limited effect on the thermal properties of the atmosphere (acting as a greenhouse gas for outgoing irradiation).  However, some long-lasting changes to the temperature structure of the atmosphere are present. For example, we find that the multi-year oscillations in atmospheric temperature (and water vapour content), which are associated with near-surface circulation cycles driven by the Earth-like orography and tidally-locked insolation (similar cycles are found on the Earth; \citealt{07055900.2022.2082914}), have been shifted out of phase by the changes induced by our cometary impact. This suggests that even individual cometary impacts, particularly massive impacts, have the potential to drive long-lasting changes in the climate.

\begin{figure*}[tbp]
\begin{centering}
\includegraphics[width=0.99\textwidth]{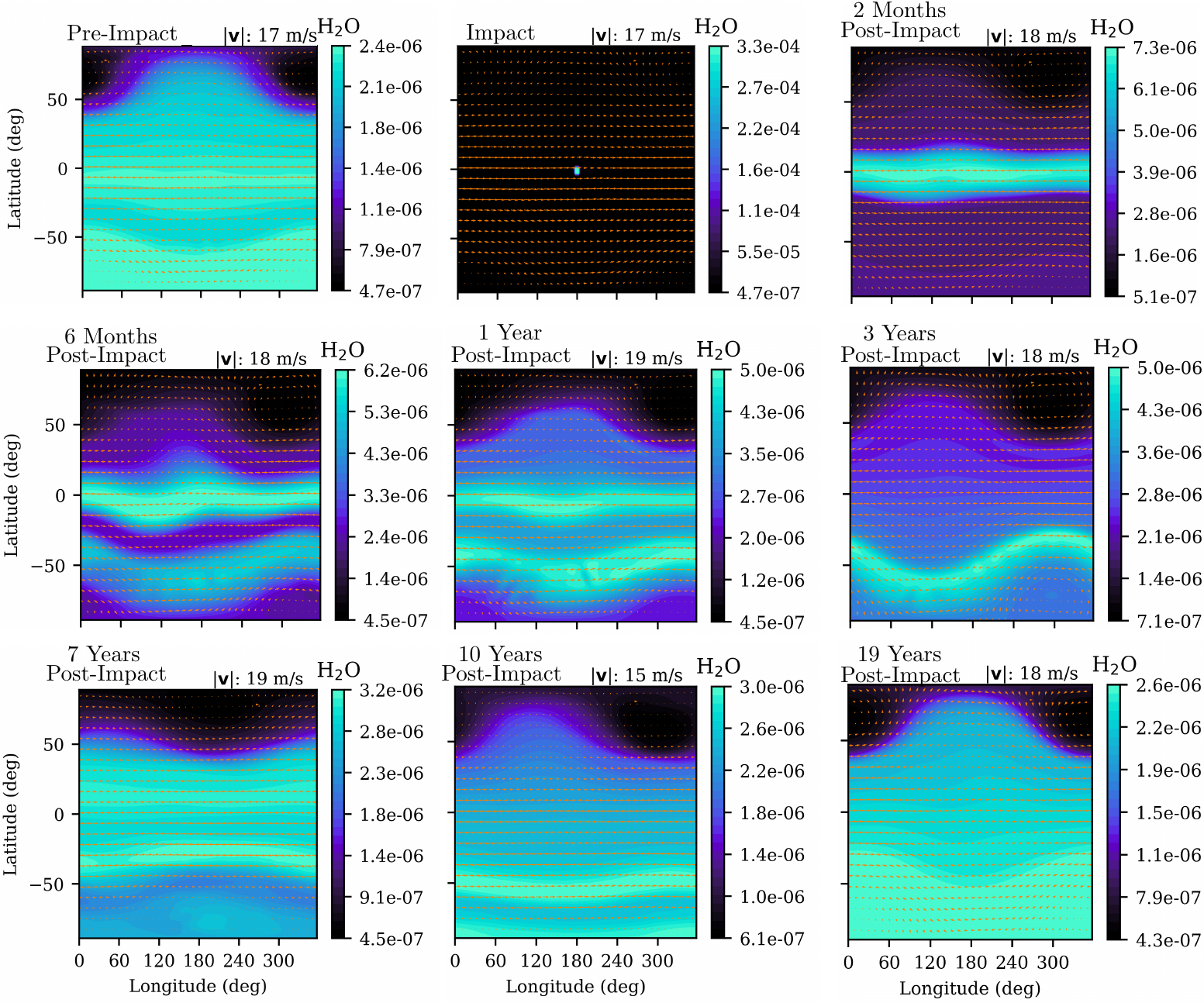}
\caption{ Horizontal slices of the fractional water abundance at a pressure of $P=2\times10^{-2}$ bar ($\sim25$ km above the surface), showing the transport of impact delivered water from the sub-stellar point both longitudinally (forming a equatorial band of water) and latitudinally (specifically towards the south-pole) {in our fiducial model}.  Note that this transport is closely associated with the horizontal wind, the mean ($|\mathbf{v}|$) of which is shown at the top right of each panel and which we plot using orange quivers. And that the dynamic range of each colour bar is different in order to highlight the change in water distribution with time.  \label{fig:water_transport_40} }
\end{centering}
\end{figure*}
\begin{figure*}[tbp] %
\begin{centering}
\includegraphics[width=0.95\textwidth]{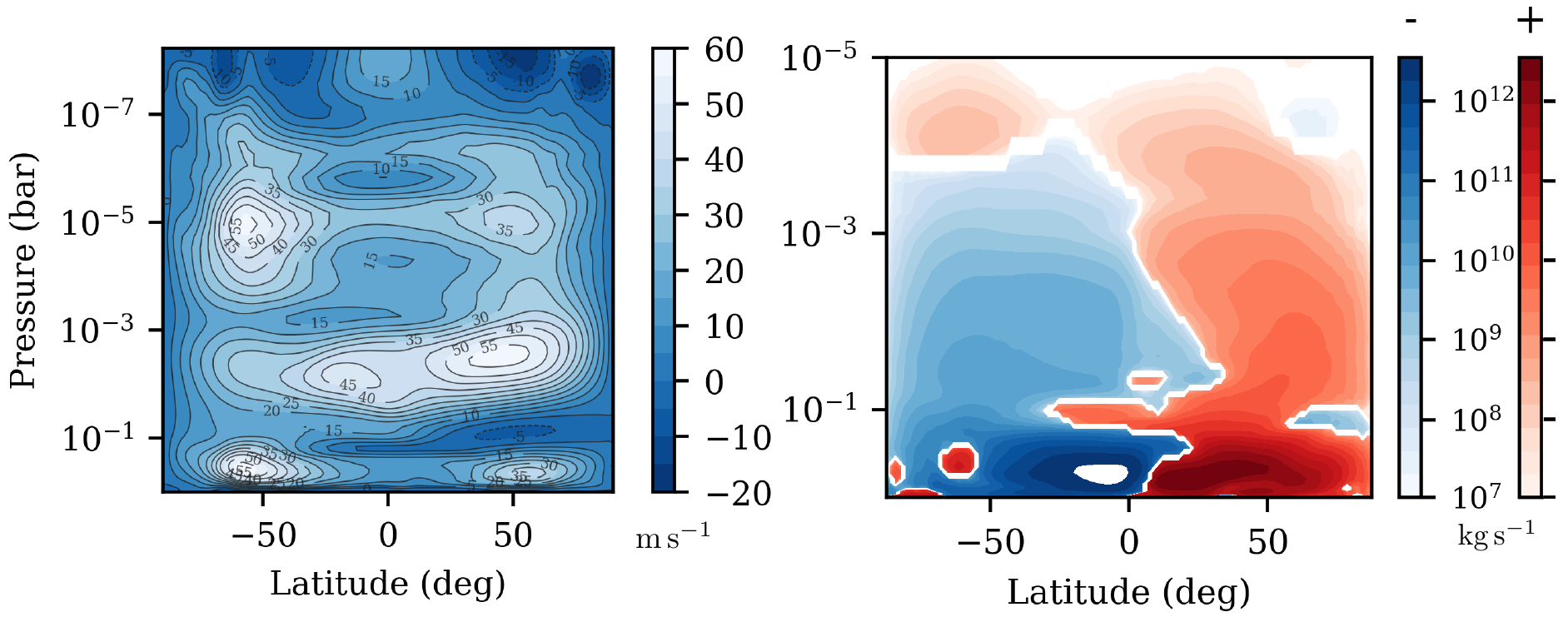}
\caption{ Zonally and temporally averaged zonal-wind (left) and meridional circulation streamfunction (right) for {our fiducial} cometary impact model. Note that the meridional circulation profile is plotted on a log scale with clockwise circulations shown in red and anti-clockwise circulations shown in blue. Thus, for example, we find that the clockwise cell in the northern hemisphere and the anti-clockwise cell in the southern hemisphere combine to drive an upflow slightly north of the equator at all pressures greater than $\sim10^{-5}$ bar. For $P<10^{-5}$ bar, which we don't show due to the relative weakness of outer atmosphere circulations, we find a series of stacked circulation cells which alternate between clockwise and anti-clockwise circulation with altitude.
\label{fig:zonal_wind_meridional_transport} }
\end{centering}
\end{figure*}
\subsection{Advection of Water} \label{sec:advection}

The post-impact enhancement in fractional water abundance at low pressures suggests that transport plays a significant role in shaping how a cometary impact can affect atmospheric composition and chemistry. This conclusion is only reinforced by the strong global-mean heating found in the mid-atmosphere, an effect which would not occur if the delivered water was confined to the impact site. As such, to better understand these effects, we next investigate the horizontal and vertical transport of water vapour. \\

We start by exploring the horizontal water vapour transport at a pressure of $2\times10^{-2}$ bar. Note that we have chosen to focus our analysis on this pressure level due to the slower dynamical timescale of the deep atmosphere, which allows us to better explore each stage of the longitudinal and latitudinal transport of water vapour. 

\autoref{fig:water_transport_40} shows the fractional water abundance at nine different points in time, ranging from pre-impact (top left), to near steady-state 19 years post impact (bottom right).
Initially water is deposited by the comet at and around the sub-stellar point (top-middle); however almost immediately the strong zonal winds at this pressure level (see \autoref{fig:zonal_wind_meridional_transport}) drive eastwards advection. As such, a little over two months post impact, we find that the impact delivered water is almost completely longitudinally homogenised. This can be seen as a strong equatorial band of water vapour in the top-right panel of \autoref{fig:water_transport_40}. Note that this horizontal homogenisation suggests that the results for cometary impacts at other equatorial longitudes should be similar to those found here, just slightly delayed (for example, the formation of the mid-atmosphere thermal inversion) to account for the advection of water from the impact site to the day-side.
Due to the inherent differences in strength between longitudinal and latitudinal transport in tidally locked atmospheres, this equatorial water band persists for months post impact. For example, the middle-left panel of \autoref{fig:water_transport_40} reveals that, six months post-impact, latitudinal winds have only just started to break the longitudinal homogenisation of the equatorial water band, with southwards/northwards advection { slightly} east/west of the sub-stellar point respectively. 
The difference in strength between longitudinal and latitudinal mixing remains apparent even as latitudinal winds advect the peak in the water abundance southwards. For example, between one and three years post impact, shown on the centre and middle-right panels of \autoref{fig:water_transport_40}, we find a that band of water vapour forms in the southern hemisphere with strong horizontal homogenisation and which traces the shape of the off-equator winds, such as the night-side polar vortices.  
From here, latitudinal transport slowly breaks this southern water band up (bottom row of \autoref{fig:water_transport_40}), such that, 10 to 20 years post-impact, we find an abundance profile which is similar, but not identical, to that found pre-impact or in our non-impacted reference case (top left). Note that the slight longitudinal shift in the abundance profile is likely linked to the delicate multi-year oscillations being out of phase with each other due to the disruptive effects of the cometary impact (\autoref{fig:comparison_time_evo_curves_water_temp}). \\
Note that the advection of the cometary impact delivered water is similar to that found at $2\times10^{-2}$ bar for most other pressure levels in our models. However there are two exceptions. The first is that at very low pressures ($P<2\times10^{-4}$ bar), the timescale of both longitudinal and latitudinal transport is short, and as such, water vapour is very rapidly mixed throughout the outer atmosphere. This has implications for observations since an increase in the opacity of these pressure levels can mask the rest of the atmosphere from spectroscopic transit observations (\autoref{sec:observational}).
On the other hand, as discussed in \citealt{sainsbury2024b}, the presence of an Earth-like land-mass distribution with its associated orography acts to drive the near-surface dynamics away from those associated with tidally-locked insolation. At the same time, evaporation of the liquid ocean at the sub-stellar point can also mask effects associated with {a sub-stellar} cometary impact since the relative change in near-surface water abundance is small. Together these effects make an analysis of the near-surface water vapour particularly tricky, and since these pressure levels are unlikely to be probed via transit spectroscopy we focus our efforts on lower pressure levels. \\

However the influence of orography on the atmospheric dynamics is still felt at lower pressures, albeit to a lesser extent than near the surface. As discussed in \citet{sainsbury2024b}, differences in the land-mass distribution, as well as the orography of said land-masses, between the northern and southern hemispheres of the Earth, and our Earth-like model, can break the symmetry between northern and southern hemisphere winds and circulations. This can be seen in both the zonally-averaged zonal-wind and meridional circulation profiles, which we plot in \autoref{fig:zonal_wind_meridional_transport}, and it can explain why our equatorial band of post-impact water is advected southwards. \\
Starting with the zonally-averaged zonal-wind, differences in both the strength of the off-equator jets as well as their latitudinal location are apparent. 
For example, the jet in the southern hemisphere at $10^{-5}$ bar is not only { $\sim50\%$} faster than its northern hemisphere counterpart, but also a few degrees further from the equator. 
Moving deeper, to $\sim5\times10^{-3}$ bar, we find that the situation is reversed. The jet in the northern hemisphere is now $\sim20\%$ faster than its southern counterpart, and the difference in latitudinal locations has grown. Whilst the jet in the northern hemisphere is still peaks around a latitude of $50^{\circ}$, the jet in the southern hemisphere is much closer to the equator (centred around $-25^{\circ}$), so close that it appears to extend across the equator and into low latitudes in the northern hemisphere. As such, any material deposited close to this pressure level, which includes a large fraction of the cometary water, will generally be associated with southern hemisphere dynamics. This can also be seen in the meridional circulation profile (see below). 
Finally, near the surface ($\sim0.5$ bar), we find a pair of high latitude jets with the jet in the southern hemisphere, which lies at a latitude which is broadly land-mass free, being on average almost twice as fast as its northern counterpart.\\

The aforementioned preference for equatorial material to be associated with southern-hemisphere dynamics is also reflected in the meridional mass streamfunction $\psi$, which takes the form:
\begin{equation}
    \psi=\frac{2\pi R_{p}}{g\cos\theta}\int^{P_{0}}_{P_{top}}v\,dP, \label{eq:meridional}
\end{equation}
where $v$ is the latitudinal velocity, $R_{p}$ is the radius of the planet, $g$ is the surface gravity, $\theta$ is the latitude, and $P_{0}$ and 
$P_{top}$ are the pressure at the surface and top of the atmosphere respectively. The right-hand panel of \autoref{fig:zonal_wind_meridional_transport} shows the zonally-averaged meridional circulation profile.
Here clockwise circulations are shown in red and anti-clockwise circulations are shown in blue. Where these circulations meet, net flows develop. For example, the clockwise circulation cell in the northern hemisphere and the anti-clockwise circulation in the southern hemisphere combine to drive a net upflow slightly north of the equator for all { $P>5\times10^{-5}$ bar}.
As for the lower pressure regions of the atmosphere (not shown here due to the significantly weaker circulation strengths at low densities), we find stacked cells alternating between clockwise and anti-clockwise circulation with altitude, hence explaining the efficient mixing of the outer atmosphere. The combination of this net upflow and efficient horizontal mixing in the outer atmosphere explains the rapid vertical mixing of deeply deposited material seen in \autoref{fig:combined_curves_water_temp}. 
The asymmetry between circulations in the northern and southern hemispheres explains the southward advection of equatorial water: at $10^{-2}$ bar an equatorial air parcels falls within the poleward (rising) region of the southern hemisphere's anti-clockwise circulation cell, as such it is rapidly advected southwards. In the same vein, material delivered by a cometary impact at slightly higher latitudes in the northern hemisphere might be expected to be advected northwards (whilst also rising). % This is an effect we intend to investigate as part of a future study. 

\subsection{Effects of Water on the Broader Atmosphere} \label{sec:water_implications}
%Photolysis
\begin{figure*}[tbp] %
\begin{centering}
\includegraphics[width=0.85\textwidth]{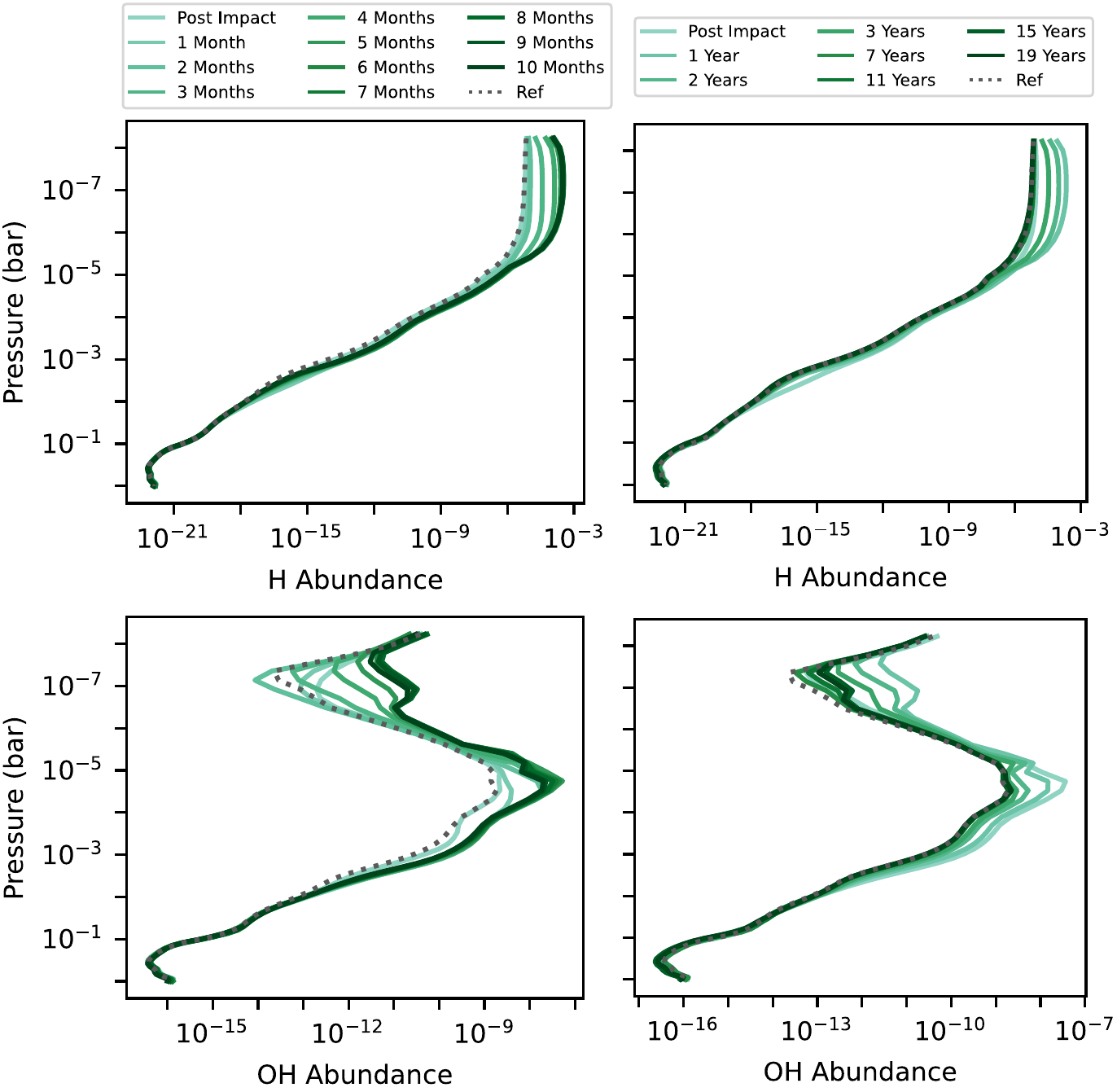}
\caption{ Fractional atomic hydrogen (H - top) and hydroxyl radical (OH - bottom) abundance profiles showing the products of the photolysis of the cometary delivered water in our {fiducial} model. On the left we show the profiles within the first 10 months of impact, when the water content of the outer atmosphere is at its peak and hence photolysis is strongest, whilst on the right we show how, over 19 years post impact, the atmosphere settles towards a quasi-steady-state close to our non-impacted reference case (grey dashed).
\label{fig:combined_photolysis} }
\end{centering}
\end{figure*}

Beyond acting as a source of heating in the mid-atmosphere, the water that an icy cometary impact delivers to the atmosphere also affects the atmospheric chemistry, composition, and climate. There are two main mechanisms by which this can occur; i) the deposited water can, via photolysis, act as a source of atmospheric oxygen and hydrogen, tipping the balance in chemical reactions towards oxygen- and hydrogen-rich molecules; and ii) the water can condense out of the atmosphere to form rain droplets, snow flakes, and ice crystals, all of which can contribute to both cloud formation and the scattering of incoming irradiation. \\

\subsection{Effects of Water on Composition}

We start by exploring the effects of the deposited water on the atmospheric composition. Underlying these changes is the photodissociation of the deposited water vapour. We investigate the enrichment of the two products which form due to one of the main UV photolysis reactions of water;
\begin{equation}
    \mathrm{H_2O} + hv \rightarrow \mathrm{H} + \mathrm{OH},
\end{equation}
atomic hydrogen (H) and the hydroxyl radical (OH). \autoref{fig:combined_photolysis} shows the change in fractional atomic hydrogen (top row) and hydroxyl radical (bottom) abundances both shortly after impact (left) and over the $\sim 20$ years required for the atmosphere to reach a quasi-steady-state (right). Here we find that both molecules exhibit a { slightly} delayed enrichment, { particularly the hydrogen abundance at low pressures.}.
This delay occurs because the initial photolysis rate of the deposited water is slow { and it takes time for the water to advect through the atmosphere}: when the deposited water is confined to the impact site the rate of photolysis is low due to the low ratio of integrated UV flux to water vapour in such a confined region. However as discussed in \autoref{sec:advection}, the deposited water is rapidly mixed zonally, and more weakly mixed latitudinally. This significantly increases the area in which UV irradiation can interact with cometary delivered water, resulting in the { $2$ to $3$} order of magnitude increase in peak atomic hydrogen and hydroxyl radical abundance seen in \autoref{fig:combined_photolysis}. 
The abundance of both molecules then drops due to a combination of further chemical reactions, such as the formation of molecular hydrogen (H$_2$ - \autoref{fig:combined_oxygenation_hydrogenation}) or hydroperoxyl radicals (HO$_2$ - \autoref{fig:combined_ozone_destruction}), and a slowing of the water photolysis rate as water vapour is advected towards to the dark night-side and poles, as well as raining/snowing/freezing out of the atmosphere. 
This drop is more rapid than the decrease in water abundance in the outer and mid atmosphere (\autoref{sec:results_combined}), suggesting that there is a non-linear relationship between water abundance and the rate of water photodissociation. 
This can be linked back to the opacity of water: when the fractional water abundance is very high ($\gtrsim10^{-5}$), the mid-atmosphere becomes optically thick, and hence most of the incoming UV irradiation is absorbed by water, leading to strong photodissociation. However, as the water abundance drops, so too does the associated opacity, and hence some of the incoming UV irradiation can be absorbed by other photosensitive molecules, such as ozone (O$_3$) or the hydroxyl radical (OH).\\ 
Note that WACCM6/CESM2 includes both a wide range of photodissociation pathways as well as a number of reaction pathways via which oxygen can be freed from hydroxyl radicals, such as the formation of hydroperoxyl radicals via the destruction of ozone - $\mathrm{OH} + \mathrm{O_3}\rightarrow \mathrm{HO_2} + \mathrm{O_2}$. For more details, see Table S2 of \citet{https://doi.org/10.1029/2019MS001882} which lists every (photo)chemical reaction included in our model. \\

%Oxygenation
\begin{figure*}[tbp] %
\begin{centering}
\includegraphics[width=0.85\textwidth]{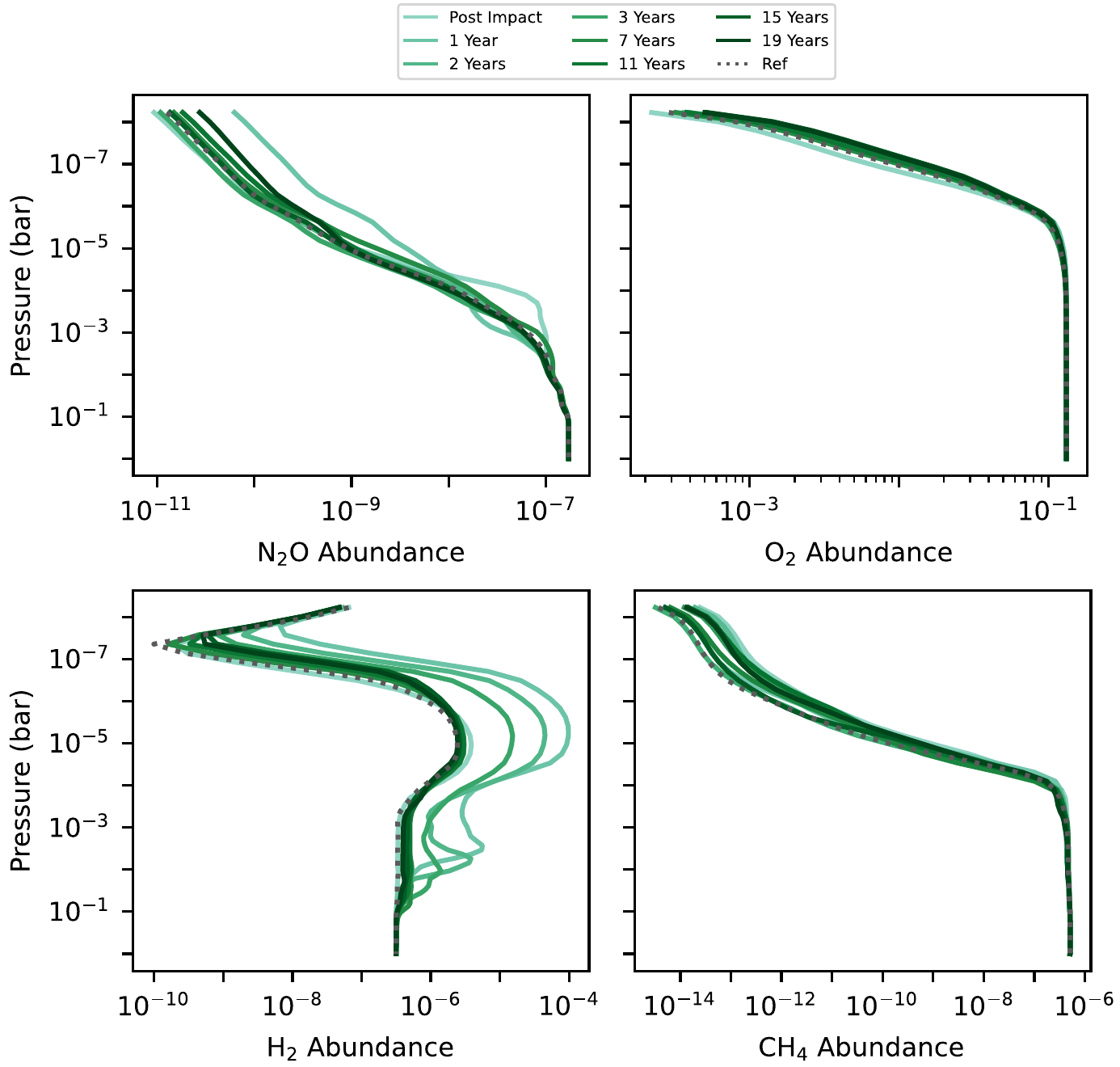}
\caption{
Fractional nitrous oxide (N$_2$O; top left), molecular oxygen (O$_2$; top right), molecular hydrogen (H$_2$; bottom left), and methane (CH$_4$; bottom right) abundance profiles for a selection of molecules whose abundance is enhanced due to the impact driven enrichment of atmospheric oxygen (top) or hydrogen (bottom). Here each profile is calculated by averaging both horizontally over all latitudes and longitudes, and temporally over a month of simulation time and we compare each profile with that found in our non-impacted reference state (grey dashed).
\label{fig:combined_oxygenation_hydrogenation}}
\end{centering}
\end{figure*}
%Ozone
\begin{figure*}[tbp] %
\begin{centering}
\includegraphics[width=0.85\textwidth]{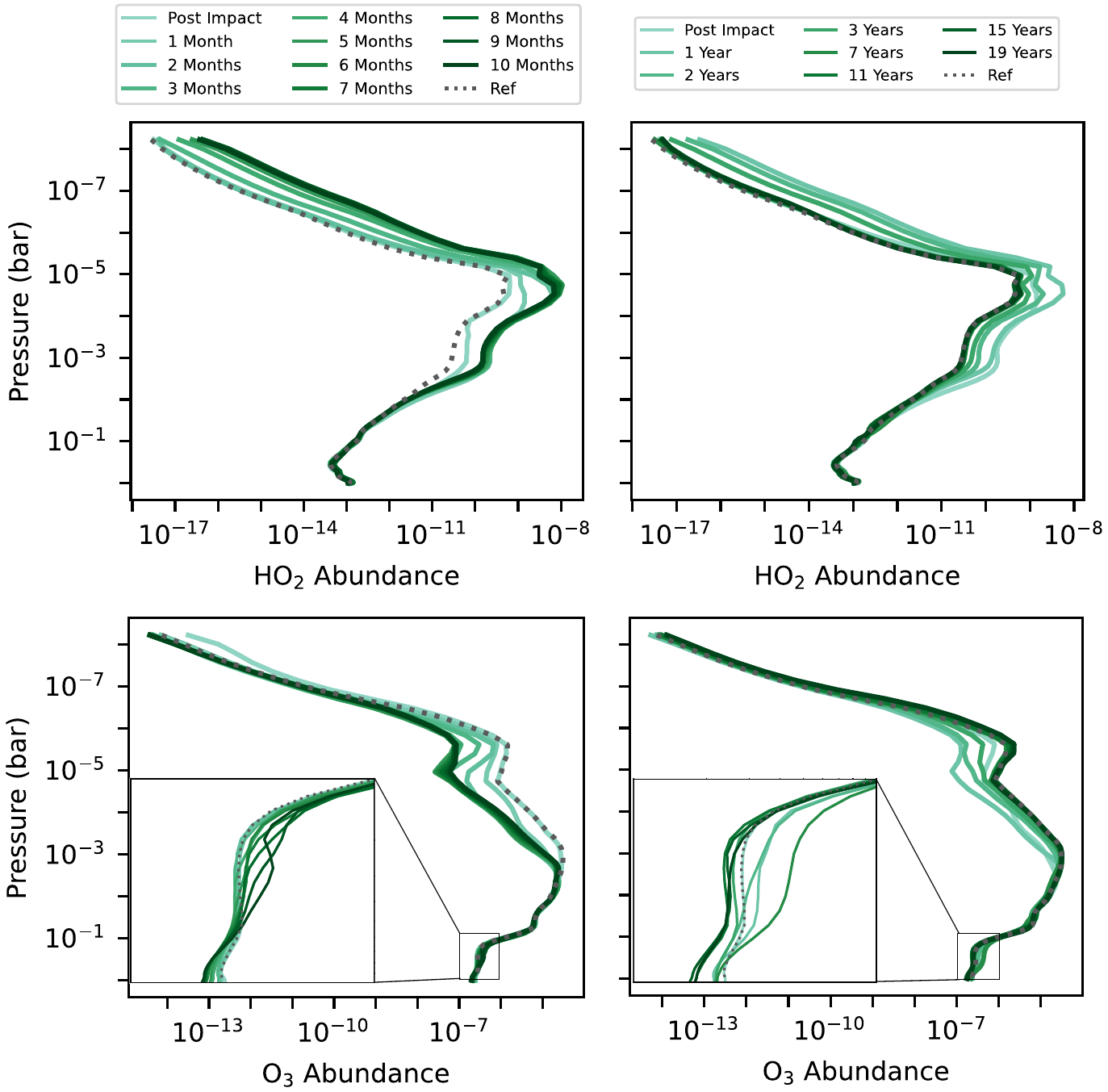}
\caption{ Fractional hydroperoxyl radical (HO$_2$; top) and ozone (O$_3$; bottom) abundance profiles showing how the increase in atmospheric oxygen abundance associated with the cometary impact does not necessarily lead to the formation of ozone. Instead we find that it actually results in a decrease in ozone abundance due to the formation of molecules associated with catalytic ozone destruction. Here we show how these molecules evolve over the first ten months post impact (left), when water photolysis is strongest, and over 19 years post impact as the atmosphere settles towards a quasi-steady-state close to that found in our non-impacted reference state (grey dashed). To better demonstrate the change in ozone abundance near the surface, we include an inset showing a zoomed-in view of the fractional ozone abundance profile between $1$ and $0.1$ bar. \label{fig:combined_ozone_destruction} }
\end{centering}
\end{figure*}

As discussed above, the photodissociation of cometary impact delivered water and the resulting products changes the overall composition of the atmosphere, enhancing the fractional abundance of oxygen-rich and hydrogen-rich molecules. Four examples of molecules which increase in abundance due to this enhancement are shown in \autoref{fig:combined_oxygenation_hydrogenation}; nitrous oxide (N$_2$O; top left), molecular oxygen (O$_2$; top right), molecular hydrogen (H$_2$; bottom left), and methane (CH$_4$; bottom right).  \\
We start with nitrous oxide (top left), which shows a strong enhancement for { $P\lesssim5\times10^{-4}$} bar one month post impact, when the photochemically driven oxygen enhancement is at its peak.
However, except at the very lowest pressures where the total number of molecules is low, we find that this enhancement does not persist. {This is again due to} photochemistry: throughout the day-side, except near the surface which is mostly shielded, we find that the incoming UV irradiation drives photodissociation of nitrous oxide, leading to the formation of molecular nitrogen (N$_2$), molecular oxygen (O$_2$), and nitrogen monoxide (NO). 
A similar result, vis a vis an initial peak in abundance followed by photo-chemical destruction, is also found for, for example, nitrogen dioxide (NO$_2$), nitrate (NO$_3$), and to a lesser extent hydroperoxyl radicals (HO$_2$; see \autoref{fig:combined_ozone_destruction}). As we discuss below, there is another mechanism by which these molecules can be destroyed: reacting with and destroying ozone. Together, these mechanisms can drive non-linear changes in fractional abundance, particularly when abundances, or the local atmospheric density, is low.   \\
We also find that the fractional abundance of molecular oxygen (O$_2$ - top right) increases due to the oxygenation of the atmosphere. However, due to the high abundance of O$_2$ (it is the second most abundant molecule in an Earth-like{/exo-Earth} atmosphere), the observed changes in abundance are both relatively small and confined to lower pressure regions where O$_2$ molecules have a lower density. As such, it is unlikely that the change in O$_2$ abundance due to an individual cometary impact would be observable, even discounting the effects that a massive increase in water vapour has on the atmospheric opacity (see \autoref{sec:observational}). {Note however that the fact that the O$_2$ abundance of the atmosphere is enhanced at all suggests that cometary impacts may be an important means of delivering oxygen to young-planets which start life in an oxygen-poor state (i.e., with a reducing atmosphere).} \\ 
We finish with two molecules which experience an enhancement due to the impact driven enrichment of atmospheric hydrogen: molecular hydrogen (H$_2$), which we plot on the bottom left of \autoref{fig:combined_oxygenation_hydrogenation}, and methane (CH$_4$), which we plot on the bottom right.
The post-impact enhancement of H$_2$ is relatively simple to understand: not only does one of the photodissociation pathways of water directly lead to the formation of H$_2$, but other photolysis products (H/OH) also undergo reactions which lead to the formation of H$_2$. As as result we find a strong enhancement in H$_2$ that persists for up to ten years post impact.\\
The enrichment in atomic hydrogen and hydroxyl radical abundance due to the cometary impact can also lead to the formation of methane via the destruction of, for example, formaldehyde (CH$_2$O). This process is reinforced by the opacity of water, reducing the UV driven photodissociation of methane. However, the limited number of reaction pathways which lead to the formation of methane, as well as the low density of the molecules involved, means that the enhancement is generally limited { and delayed with respect to the impact}.\\ 

There is one {major} exception to the impact-driven enhancement in the abundances of oxygen bearing molecules: ozone (O$_3$). Rather than being enhanced, we instead find significant ozone depletion in the mid atmosphere, especially around $10^{-5}$ bar. This can be seen in \autoref{fig:combined_ozone_destruction}, which shows the change in the fractional abundance of ozone (bottom row) and hydroperoxyl radical (top row) both shortly after impact (left) and over the $\sim20$ years required for the atmosphere to reach a quasi-steady-state (right).\\
Here, as shown on the bottom row of \autoref{fig:combined_ozone_destruction}, we find a { two} order of magnitude depletion in ozone abundance { three to four} months post impact.
Furthermore, even though this depletion is primarily in the mid-atmosphere, and our models reveal an {initial}, weak, enhancement in near-surface ($P>10^{-1}$ bar) ozone post-impact, we find that, for the first five years post-impact, the average ozone column density has dropped by { $\sim7.5\%$} from $\sim7.7\times10^{23}$ molecules\,m$^{-2}$ to { $\sim7.1\times10^{23}$} molecules\,m$^{-2}$. This destruction, as well as the slight post-impact deep enhancement, can be linked to the cometary delivery of water.\\
As discussed above, the oxygenation of the atmosphere by cometary material leads to the formation of NO$_\mathrm{x}$ and HO$_{\mathrm{x}}$, families of molecules which play a key role in the destruction of ozone on the Earth. For example, hydroperoxyl radicals (HO$_{2}$), which are thought to be responsible for around half of ozone destruction in the Earths atmosphere \citep{doi:10.1126/science.266.5184.398}, exhibit a significant increase in abundance, over an order of magnitude, throughout the mid and outer atmosphere ($P<10^{-3}$ bar).
This enrichment occurs because there are numerous reaction routes via which these molecules can form, including pathways which result in the destruction of ozone, such as  
\begin{equation}
\mathrm{O_3}+\mathrm{OH}\rightarrow\mathrm{HO_2}+\mathrm{O_2}.
\end{equation}
The hydroperoxyl radicals which forms from this pathway can also destroy ozone,
\begin{equation}
\mathrm{O_3}+\mathrm{HO_2}\rightarrow\mathrm{OH}+\mathrm{2O_2},
\end{equation}
leading to a chain reaction, including a catalytic hydroxyl radial driven ozone destruction cycle, and the strong correlation between the hydroperoxyl radical enhancement and ozone depletion shown in \autoref{fig:combined_ozone_destruction}.
This cycle of NO$_\mathrm{x}$/HO$_\mathrm{x}$ formation and ozone depletion weakens as the water photolysis rate drops (\autoref{fig:combined_photolysis}), until approximately { seven} years post impact where we find a fractional ozone abundance that is similar to our non-impacted reference state. Note, however, that there is an exception to this at very low pressures where we instead find a weak but persistent enhancement. This enhancement is likely associated with the slight increase in overall oxygen content of the atmosphere adjusting the balance between ozone destruction and formation. \\
{Similarly,} mid-atmosphere water photolysis is also responsible for the slight enhancement in near-surface ozone between six months and { three} years post impact. This is because the deposited water acts as a both a strong source of opacity, shielding lower altitudes from the incoming (UV) irradiation, thus shielding deep ozone from photodissociation. {As well as a source of free O, which can react with O$_2$ to form ozone.} However, as the water content of the mid-atmosphere drops, so too does its opacity, leading to the UV irradiation once again penetrating into the deep atmosphere and a steady { drop in the near surface ozone abundance}. \\

\begin{figure*}[tbp]
\begin{centering}
\includegraphics[width=0.85\textwidth]{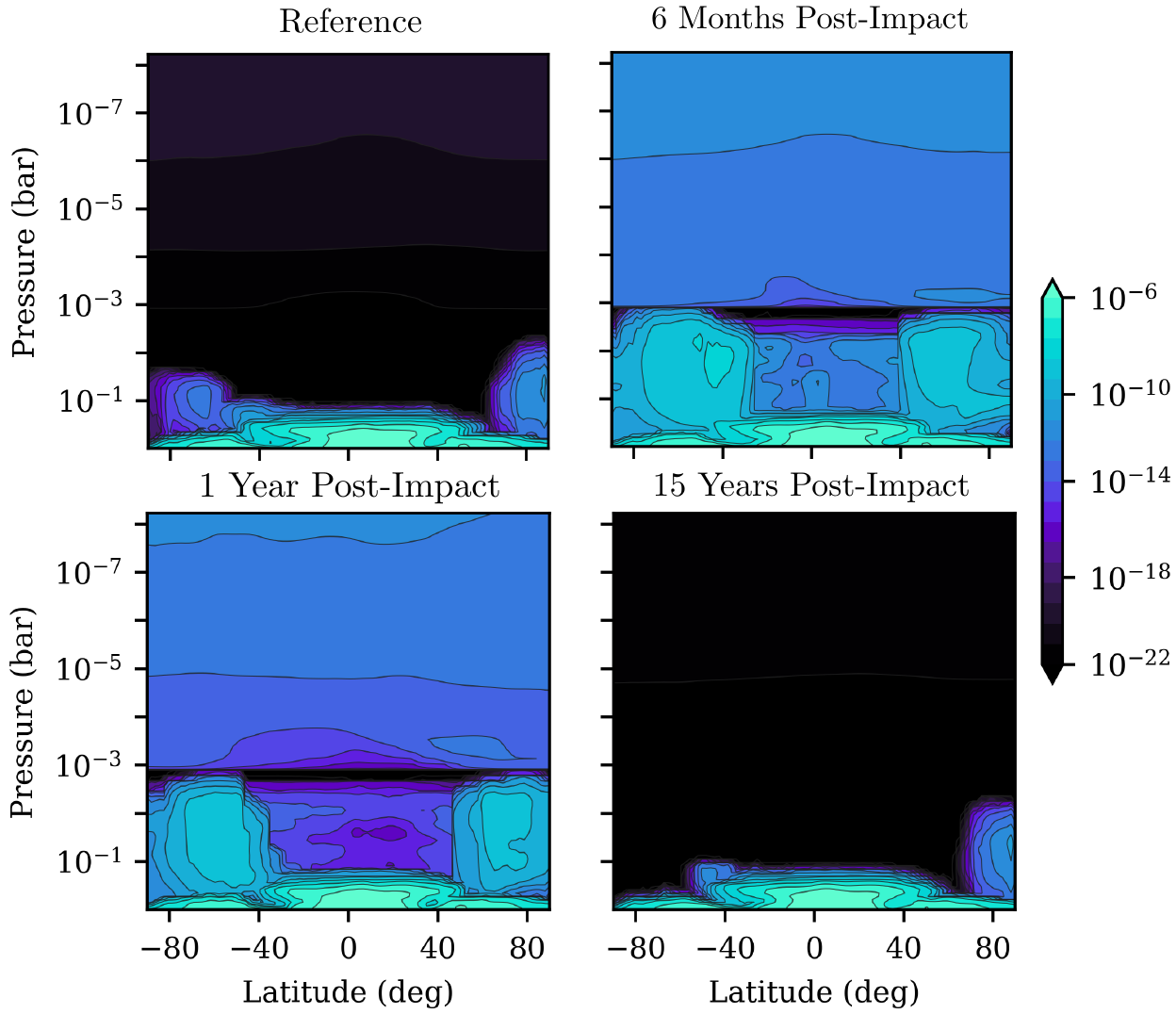}
\caption{ Zonally and temporally averaged snow fraction for our non-impacted reference model (top left) and at three points in time post-impact for our {fiducial} model: 6 months (top right), 1 year (bottom left), and 15 years (bottom right) post-impact, by which time the impacted models snow fraction has returned to a state close to that found in our reference state.   \label{fig:combined_snow_fraction_MR} }
\end{centering}
\end{figure*}

\begin{figure*}[tbp]
\begin{centering}
\includegraphics[width=0.85\textwidth]{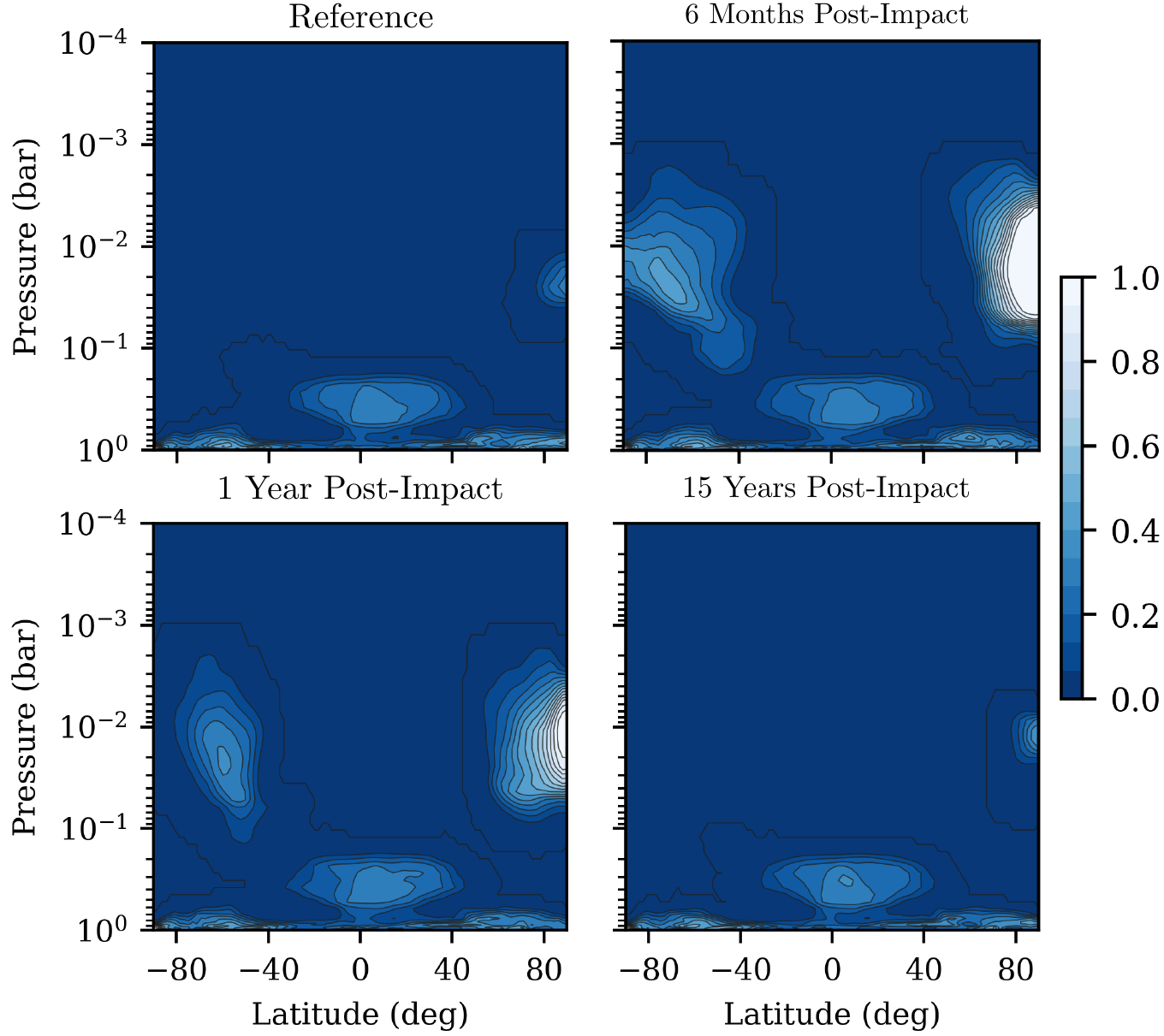}
\caption{ Zonally and temporally averaged cloud fraction for both our non-impacted reference state (top left) and at three points in time post-impact for our {fiducial} model: 6 months (top right), 1 year (bottom left), and 15 years (bottom right) post-impact, by which time the impacted models cloud fraction has returned to that found in our non-impacted reference state. Here we limit our plots to pressures $>10^{-4}$ bar due to the limited pressure range (P$>10^{-3}$ bar) that CESMs cloud model operates over.    \label{fig:combined_cloud} }
\end{centering}
\end{figure*}

\subsection{Effects of Water on Climate}

As an Earth-system model, WACCM6/CESM2 includes a robust treatment of not only the atmospheric chemistry and dynamics, but also the climate, including cloud formation and precipitation, i.e, rain and snow. Since, for an Earth-like atmospheric composition, the primary constituent of both is water, it can be inferred that they will be significantly affected by an icy cometary impact. \\
In \autoref{fig:combined_snow_fraction_MR} and \autoref{fig:combined_cloud} we plot the zonally and temporally averaged snow fraction, which represents the fraction of each cell's mass which is made up of snow, and cloud fraction, which represents the fraction of each cell that is covered by clouds, respectively, at three different points in time post-impact, comparing our profiles with our non-impacted reference state. 
In both cases we find significant changes associated with the cometary impact delivered water. These changes are most visually apparent in the snow fraction profile, where we find a massive increase, greater than ten orders of magnitude, in snow fraction at almost all pressure levels $<0.1$ bar. This is significant enough that low-pressure snow (and ice) now makes up a measurable fraction of the outer and mid atmosphere, unlike the reference case where the presence of similar snow fractions is limited to near-surface regions.
As we discuss in \autoref{sec:observational}, this enhancement in low pressure snow, and cloud ice particles, can have significant implications for the observed transmission spectra. \\
The only exception to the above increase in snow fraction is found for pressures slightly greater than $10^{-3}$ bar, where we instead find a strong dip associated with cloud formation and the shift of atmospheric water from snow/ice to clouds. 
Note, however, that the dip which occurs here is somewhat artificial: in its current form, WACCM6/CESM2 only models clouds for pressures $>10^{-3}$ bar. 
Cloud formation also explains the differences in the time evolution of the snow fraction above and below this level, with the atmosphere returning towards the non-impacted reference state faster for pressures $>10^{-3}$ bar. \\
Evidence for cloud formation between $10^{-1}$ and $10^{-3}$ bar can be seen in the cloud fraction profile (\autoref{fig:combined_cloud}). Here we find significant cloud formation near the poles, with the cloud fraction reaching one between { $\sim10^{-1}$ and $\sim10^{-3}$} bar at the north-pole.
These high-altitude, off-equator/polar, clouds form due to the condensation of impact-delivered water vapour in polar vortices and cold night-side Rossby gyres, the latter of which are similar to those found by \citet{10.1093/mnras/stad2704}. However nearer the surface, $P>10^{-1}$ bar, we find that orography breaks up these water vapour confining circulations \citep{sainsbury2024b}, and instead the main driver of cloud formation is ocean-evaporation at the sub-stellar point. Hence the equatorial cloud patch found at $\sim0.5$ bar. \\
Since the formation of high-altitude clouds is so closely tied to the water vapour enrichment of the mid-atmosphere, we find that, as this enhancement drops, so too does the high-latitude cloud fraction. Whilst the drop seen in the first year post-impact is fairly rapid, it takes approximately 15 years for the cloud fraction profile in our fiducial model to return to a state reminiscent of our non-impacted reference state (bottom right). \\
It is important to note that, whilst cloud formation might be expected to have an effect on the planetary albedo by increasing the fraction of the incoming irradiation which is reflected, the high-latitude and/or night-side formation location of these impact induced clouds means that any effect will be limited. In fact, one might expect that the clouds would instead cause the surface to warm due to their greenhouse effect, and indeed an analysis of the long-wave cloud forcing, which measures the greenhouse effect of clouds due to absorption and re-emission of outgoing radiation, suggests that this is the case. For example, in the first six months post-impact, we find a $\sim10\%$ increase in the long-wave cloud forcing. However the observed cooling of the deep atmosphere post-impact suggests that the drop in near-surface heating due to high-altitude water vapour absorbing incoming irradiation on the day-side counters this warming greenhouse effect.

%PSG transmission spectra
\begin{figure*}[tbp] %
\begin{centering}
\includegraphics[width=0.99\textwidth]{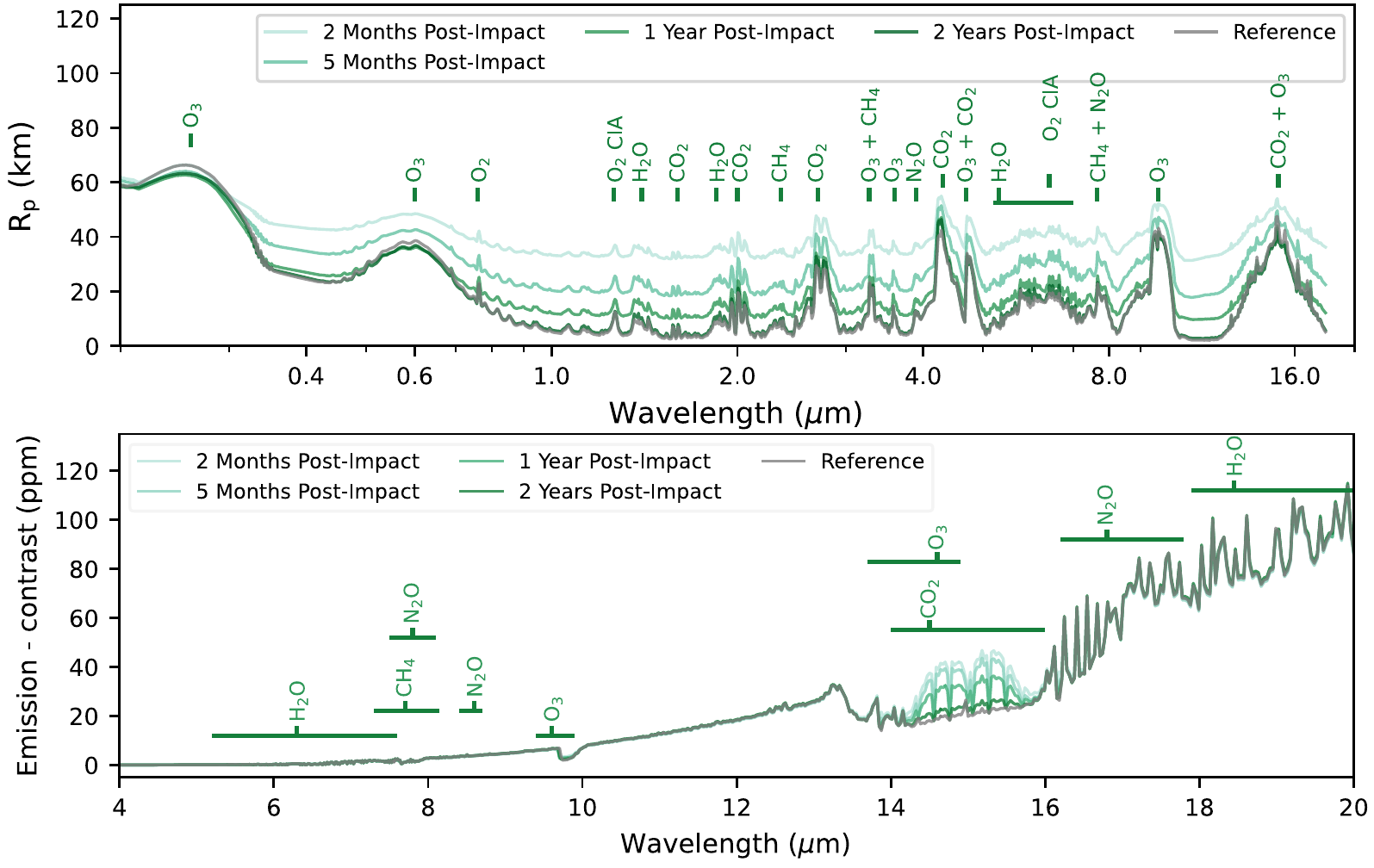}
\caption{ Example transmission spectra (top), in units of transit atmospheric thickness (R$_p$), and emission spectra, in units of contrast (radiance ratio with respect to the host star in parts per million), calculated with PSG using 3D data from {fiducial} cometary impact model at four different points in time post impact: two months, five months, one year, and two years. We also include a reference spectra, in grey, calculated using PSG using data from our non-impacted reference state. 
To aid in interpretation, we have labelled a number of spectroscopic features of interest, including lines associated with H$_2$O, CO$_2$, CH$_4$, N$_2$O, O$_2$, and O$_3$. The emission spectra are calculated at a phase of 90$^\circ$, i.e. midway between primary and secondary eclipse, which was chosen to be representative of the `average' case/emission. Note that we do not include any spectra calculated at earlier times in our impacted models due to instabilities in PSG caused by scattering associated with the high water content of the very outer atmosphere. 
\label{fig:time_evolution_transmission_emission}  }
\end{centering}
\end{figure*}
\begin{figure*}[tbp]
\begin{centering}
\includegraphics[width=0.85\textwidth]{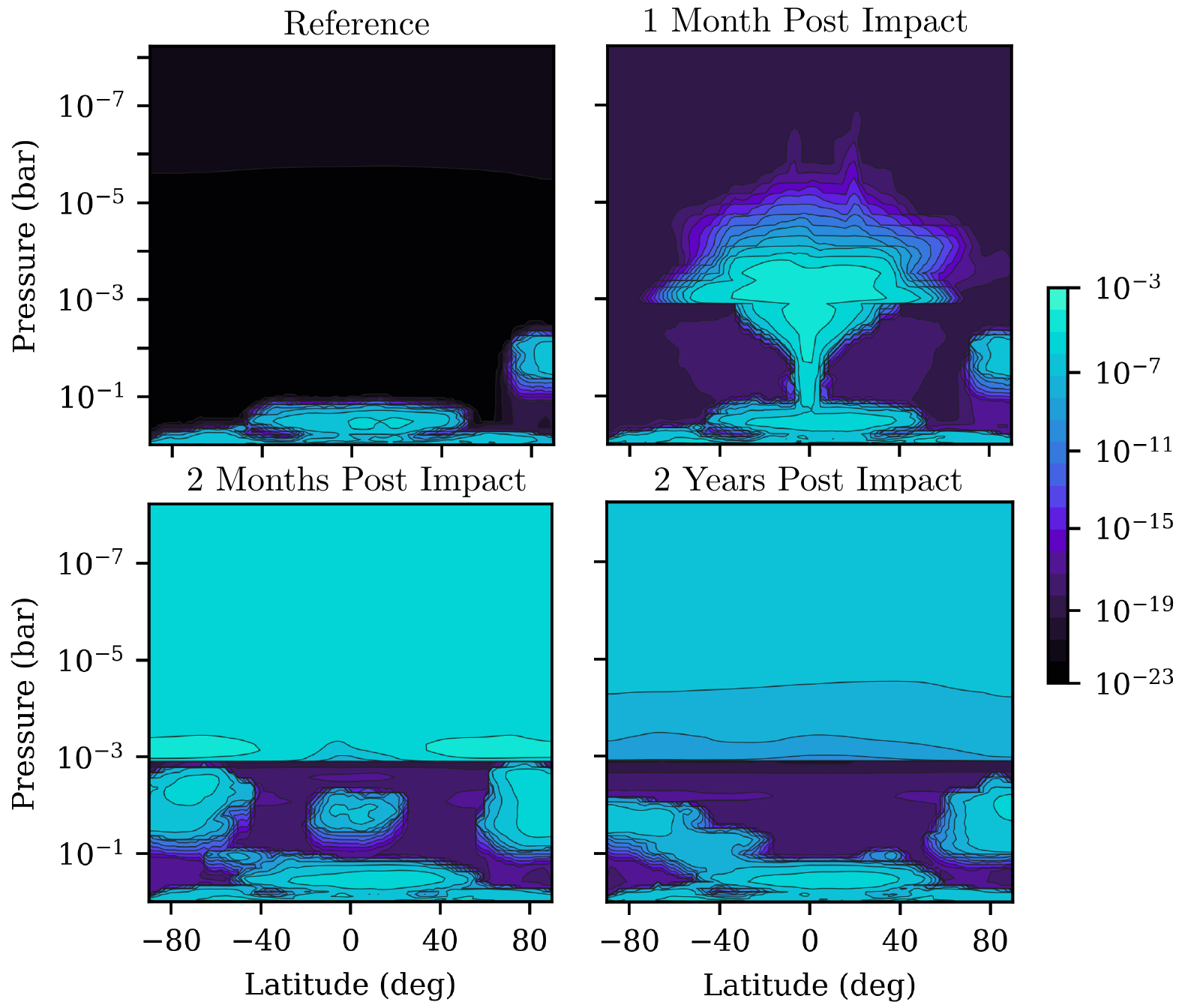}
\caption{ Zonally and temporally averaged cloud ice mixing-ratio for our reference non-impacted model (top left) and at three points in time {for our fiducial} model: 1 month (top right), 2 months (bottom left), and 2 years (bottom right) post-impact, with the latter times selected to correspond to the spectra shown in \autoref{fig:time_evolution_transmission_emission}.   \label{fig:combined_cloud_ice_MR} }
\end{centering}
\end{figure*}
\subsection{Observational Implications} \label{sec:observational}

We finish by investigating if the changes in atmospheric chemistry, composition, and climate that result from the impact of a single pure water ice comet (with $R=2.5$ km and $\rho=1\,\mathrm{g cm^{-3}}$) with a tidally-locked, Earth-like, exoplanet, might be observable. 
To do this we use the Planetary Spectrum Generator (PSG - \citealt{2018JQSRT.217...86V}) to calculate idealised transmission and thermal emission spectra at four points in time post-impact, comparing these spectra with a corresponding spectra calculated using our non-impacted reference state. We use the GlobES 3D mapping tool to compute transmission and thermal emission spectra between $0.2$ and $20\,\mathrm{\mu m}$ using three-dimensional data from a five-day temporal-average snapshot. These spectra are binned both spectrally, such that spectral resolution is $R=250$, and spatially. This is necessary to both reduce the file size given to PSG as well as the computational resources required to calculate a transmission or emission spectrum from 3D data. We regrid the data used to calculate the spectra to a $10^{\circ}$ resolution longitudinally whilst leaving the latitudinal resolution unchanged (as this has the largest effect on transmission spectra).  
Note that we do not plot the emission spectrum between $0.2$ and $4\,\mathrm{\mu m}$ due to a lack of any significant emission in this region. The thermal emission for TRAPPIST-1e only becomes significant for wavelengths $>8\mathrm{\mu m}$ \citep{2019arXiv190802356S} due to the planets low equilibrium temperature ($\sim240$ K). \\
The resulting transmission and thermal emission spectra are shown in \autoref{fig:time_evolution_transmission_emission}. Here we plot spectra at four points in time post-impact (green), comparing them with a corresponding spectra calculated from our non-impacted reference case (grey). 
To aid in our analysis, we also include labels indicating a number of spectroscopic features of interest such as those associated with water (H$_2$O), carbon dioxide (CO$_2$), methane (CH$_4$), nitrous oxide (N$_2$O), oxygen (O$_2$), and ozone (O$_3$). \\

We start with the transmission spectra calculated two months post impact, after the deposited water has started to mix throughout the outer atmosphere. Here we find a near uniform {(for wavelengths $>0.8\,\mathrm{\mu m}$)} increase in the apparent thickness of the atmosphere { ($\sim\,38$ km)} when compared to our non-impacted reference state ($\sim14$ km). This suggests that our transmission spectra is probing the lowest pressure regions of our atmosphere, a conclusion which is reinforced by the strong suppression of almost all spectral features, { a suppression which occurs} because the densities of many absorbers in the probed region are low. But why are our transmission spectra probing such a low density region post-impact? The answer lies in the formation of snow (\autoref{fig:combined_snow_fraction_MR}) and cloud ice. The latter is shown in \autoref{fig:combined_cloud_ice_MR}, where we plot the zonally-averaged cloud ice mixing ratio at three different points in time post-impact, comparing these profiles with our non-impacted reference case. Here we find that the water delivery associated with an icy cometary impact drives the formation of significant quantities of cloud ice at pressures $<10^{-3}$ bar. Specifically, we find that the fraction of the atmosphere that is made up of cloud ice has gone from being insignificant to a few tenths of a percent. And whilst this still represents a small fraction of the atmosphere, it is enough to scatter a significant fraction of the incoming radiation from the host-star, leading to the observed increase in `continuum' level, and hence apparent atmospheric thickness, in our transit spectra. {Note that the scattering driven increase in the apparent radius of a planet post a recent impact should be identifiable from a single transit observation when comparing the apparent radius with pre-impact or follow-up observations. }
\\
With time, this cloud ice slowly settles, evaporates, and/or rains and snows out, decreasing the strength of this scattering effect and allowing us to probe lower-altitude, denser, regions of the atmosphere. For example, five months post impact, we find that between $1$ and $18\,\mathrm{\mu m}$, our transmission spectrum now probes an average altitude of { $\sim\,28$} km above the surface, and spectral features associated with all of the aforementioned species are { stronger}, albeit still somewhat suppressed, { relative to our reference state,} due to the relatively low densities probed. \\
One year post-impact, between $1$ and $18\,\mathrm{\mu m}$, we find that the transmission spectrum now probes an average altitude of { $\sim\,19$} km, which is only slightly higher than our non-impacted reference state ($\sim\,14$ km), allowing for compositionally driven differences in spectroscopic features to become { more} apparent. 
For example, we tend to find that the strengths of water features, such as those found at $1.4\,\mathrm{\mu m}$, { $1.9\,\mathrm{\mu m}$,} or between $5.2$ and $7\,\mathrm{\mu m}$, are similar between our one year (and five month) post-impact and reference models. This suggests that, in our impacted model, the enhancement in fractional water abundance balances the drop in probed atmospheric density. 
On the other hand, the strengths of ozone (O$_3$) features, such as those at $0.6\,\mathrm{\mu m}$ or $9.6\,\mathrm{\mu m}$, are clearly reduced, { with the former actually being weaker than our reference case and the latter} only resulting in slightly higher apparent radii than the non-impacted reference state, despite the still significant differences in continuum level.
A similar story, { of a slight suppression of feature strength,} holds true for other absorbers, likely due to the slightly lower densities probed, albeit to a lesser extent as many of these molecules exhibit enhanced low-pressure abundances as opposed to the reduction found for ozone (see \autoref{sec:water_implications}). {Note that, at this point in time post-impact, impact-driven changes in the strength of spectroscopic features are relatively slow to evolve, and hence should remain observable when combining transit observations taken over a period of weeks to months (as is typically done for habitable zone planets around M-dwarfs), assuming that activity from the host M-star does not wash out the features entirely (see the discussion of this problem by, for example, \citealt{2024arXiv240312617D}.}  \\
Finally, two years post-impact we find a transmission spectrum which is extremely similar to our non-impacted reference state, albeit with some slight differences in feature strength, particularly ozone, { although other oxygen-rich molecules also show slight changes, such as an enhancement in the strength of CO$_2$ features}. Unfortunately most if not all of these differences are unlikely to be distinguishable observationally given that, even when we combine multiple transits, they fall below the noise floor of JWST (see, for example, the noise estimates of \citealt{Rustamkulov_2022}). \\

The effects of a single icy cometary impact on the thermal emission spectrum, which we show on the bottom row of \autoref{fig:time_evolution_transmission_emission}, are more muted. Primarily this is because cloud ice scattering does not have the same effect on planetary emission as it does on a transmission spectrum. As such, at most of the wavelengths considered here, we find spectral feature differences which are far below the noise floor of JWST or other upcoming space-based telescopes. The only exception to this is between $\sim\,14$ and $\sim\,16\,\mathrm{\mu m}$, where we find a water-opacity driven change in the CO$_2$ feature. Specifically, we find that the increased opacity of the atmosphere masks CO$_2$ absorption features leading to an emission spectrum that is closer to a blackbody. When comparing this region with our non-impacted reference state, we find a $\sim25$ ppm enhancement two months post impact, and the change remains potentially observable for up to a year post impact (with a { $\sim\,15$} ppm enhancement). Note that a similar, { but much weaker and hence likely unobservable,} effect is found for the ozone feature at $\sim\,9.8\,\mathrm{\mu m}$, where we again find a slight ({ a few} ppm) increase in thermal emission contrast post-impact. Here however, the difference is not driven by water opacity masking the ozone feature but instead it is due to a decrease in ozone abundance. This reinforces the need to take care when interpreting emission and transmission spectra due to the complex, and wavelength dependent, relationships between observed abundances and the strength of atmospheric features. \\

Given all of the above as well as an understanding of current, and near future, observational capabilities, we suggest that individual cometary impacts are likely to be detectable only for a short period post-impact. In the transmission spectrum, this will be driven by the effects of cloud ice scattering at low pressures leading to a {substantial and measurable} increase/decrease in apparent planetary radius and spectroscopic feature strength, respectively. Atmospheric water plays a similar role on the emission spectra, masking atmospheric features in a way that may be observable for a short while post-impact. \\
One way this might manifest is as a difference between repeated observations of the same object, an effect which might indicate that a cometary impact has changed the observed atmospheric chemistry/composition for one (or more) of our observations. As the number of atmospheres characterised increases, and we take repeat observations of these atmospheres in order to further probe their composition, the chances that such a scenario will occur increases. \\
Beyond the short-lived effects associated with a cometary impact, repeated impacts, or ongoing bombardment, have the potential to deliver enough material to drive a global change in atmospheric chemistry, composition, and climate (as {might have been the case for the early} Earth).

\section{Concluding Remarks} \label{sec:concluding} 

In this work we have coupled the cometary impact model of \citet{2024ApJ...966...39S}, which includes thermal ablation at low pressures and break-up deeper with the atmosphere, with the Earth-system model WACCM6/CESM2. This coupled model was then used to simulate the impact of a pure water ice comet (with $R=2.5\,\mathrm{km}$ and $\rho=1\,\mathrm{g\, cm^{-3}}$) with the day-side {(sub-stellar point)} of a tidally-locked, terrestrial, exoplanet modelled on the potentially habitable exoplanet TRAPPIST-1e. {Our fiducial model included both the mass (water) and thermal energy deposition associated with the cometary impact, and we compared our results with a non-impacted reference state.} \\
{Our analysis of this fiducial coupled model, as well as two additional models (\autoref{sec:results_isolated}) which} isolated the effects of cometary water and thermal energy delivery, thus allowing us to investigate the strength, and timescale, of the atmosphere's response to each component of the impact, {and revealed the significant role that water delivery can play in modifying our exoplanetary atmosphere. The main takeaway results of our study are as follows:

\begin{itemize}
    \item Whilst the majority of the cometary water/ice is deposited at pressures $>10^{-5}$ bar, vertical advection  then rapidly carries this water aloft, enriching the atmospheric abundance for all pressures, including a multi-order-of-magnitude enhancement in the outer atmosphere.
    \item This enhancement in mid- and outer-atmosphere water persists for at least ten-years post impact.
    \item The impact of the cometary heating is more limited: it increases the fractional water abundance at { higher}-pressures by evaporating/sublimating atmospheric water droplets (e.g. rain/snow) and ice particles. 
    \item Deposited water acts strong opacity source, driving localised day-side heating (an increase of up to { 35} K) that peaks in strength at $\sim5\times10^{-4}$ bar. In turn, this enhanced mid-atmosphere opacity reduces the surface insolation, reducing the (near)-surface temperature by upto $2$ K.
    \item The deposited water, due to its high UV opacity, rapidly photodissociates, increasing the abundance of hydrogen and hydroxyl radicals (OH) by up to $\sim3$ orders of magnitude. 
    \item The formation and further photodissociation of hydrogen and hydroxyl radicals (OH) leads to an increase in the hydrogen and oxygen content of the atmosphere, and the formation of, for example, H$_2$, O$_2$, CH$_4$, N$_2$O, NO, NO$_2$, NO$_3$, and HO$_2$. 
    \item Despite the increase in atmospheric oxygen, the formation of NO$_\mathrm{x}$, and HO$_\mathrm{x}$ result in significant ozone (O$_3$) destruction. For example, the column integrated ozone content drops by { $\sim7.5\%$} during the first five years post-impact.
    \item Zonal winds, driven by the tidally-locked insolation, rapidly homogenise the deposited water zonally, suggesting that similar results to those found here should apply for any equatorial impact. 
    \item Horizontal winds also influence the water photolysis rate, enhancing it at early times by increasing the surface area of enriched water interacting with UV irradiation, and decreasing it at later times by advecting water to the dark night-side. Such an effect can only be captured when considering the full-3D atmospheric circulation.
    \item The deposited water drives the formation of high-latitude clouds and high-altitude cloud-ice. Scattering by this cloud-ice significantly affects the transmission spectra post-impact, increasing the apparent near-infrared atmospheric thickness from $\sim\,14$ km to { $\sim\,38$} km two months post-impact, reducing to a { $\sim\,5$} km enhancement one year post-impact.
    \item The impact-induced changes in atmospheric composition also affect the strength of absorption and emission features in the transmission and emission spectra. For example, we find a decrease in the strength of ozone absorption and emission features, and an increase in the strength of water absorption features. We also find that the increased opacity of the atmosphere masks deep CO$_2$ features in the post-impact emission spectrum. 
\end{itemize}}

{The aforementioned results vis a vis the transmission and thermal emission spectra suggest that the effects of a cometary impact should be most visible within a year of the impact, with the strong-cloud ice driven changes to the transmission spectrum being potentially detectable from a single transit observation. However, as the atmospheres settles back towards the non-impacted reference state, we find that the long-lasting changes associated with a single cometary impact fall below the noise floor of modern near-infrared telescopes, making them unlikely to be observed. As such, the most-likely scenario under which a single large impact will be observed is a short-lived (i.e. single observation) change in the spectra of a planet that has undergone repeated observations. However we acknowledge that such an event is rather unlikely for a terrestrial planet with a potentially habitable secondary atmosphere. For example, \citet{10.1046/j.1365-8711.2000.03568.x}, calculate that the time interval between impacts which lead to significant, $>2$ km diameter, craters should be more than 190,000 years. On the other hand, given the massive numbers of planets that have been, and soon will be detected, we also expect that ongoing monitoring of these objects may be able to reveal such an event.\\

Instead, the fact that impact driven changes in atmospheric chemistry and composition persist to quasi-steady-state suggests another scenario. It is possible that repeated or ongoing bombardment might drive large-scale and long-term changes which might even play a role in shaping planetary habitability. This is particularly true for young planets, where we expect the bombardment rate to be significantly higher (as was the case for the Earth - e.g. \citealt{Fassett2013,2020AsBio..20.1121O}), and for which icy comets may play an important role in delivering volatile materials (see, for example, \citealt{OWEN1995215,2002ESASP.518....9E,2018SoSyR..52..392M,2023PhyU...66....2M}). }
We will explore this in a future study, using our coupled impact/climate model to study the effects of repeated cometary bombardment {for both oxygen-rich and oxygen-poor, nitrogen dominated, atmospheres. However in the next paper in this series we will compare the results from this study to a model of an icy cometary impact with a true exo-Earth analogue planet that includes a diurnal cycle, and hence rather different atmospheric circulations and water insolation patterns} (Paper II).

\begin{figure*}[tbp]
\begin{centering}
\includegraphics[width=0.85\textwidth]{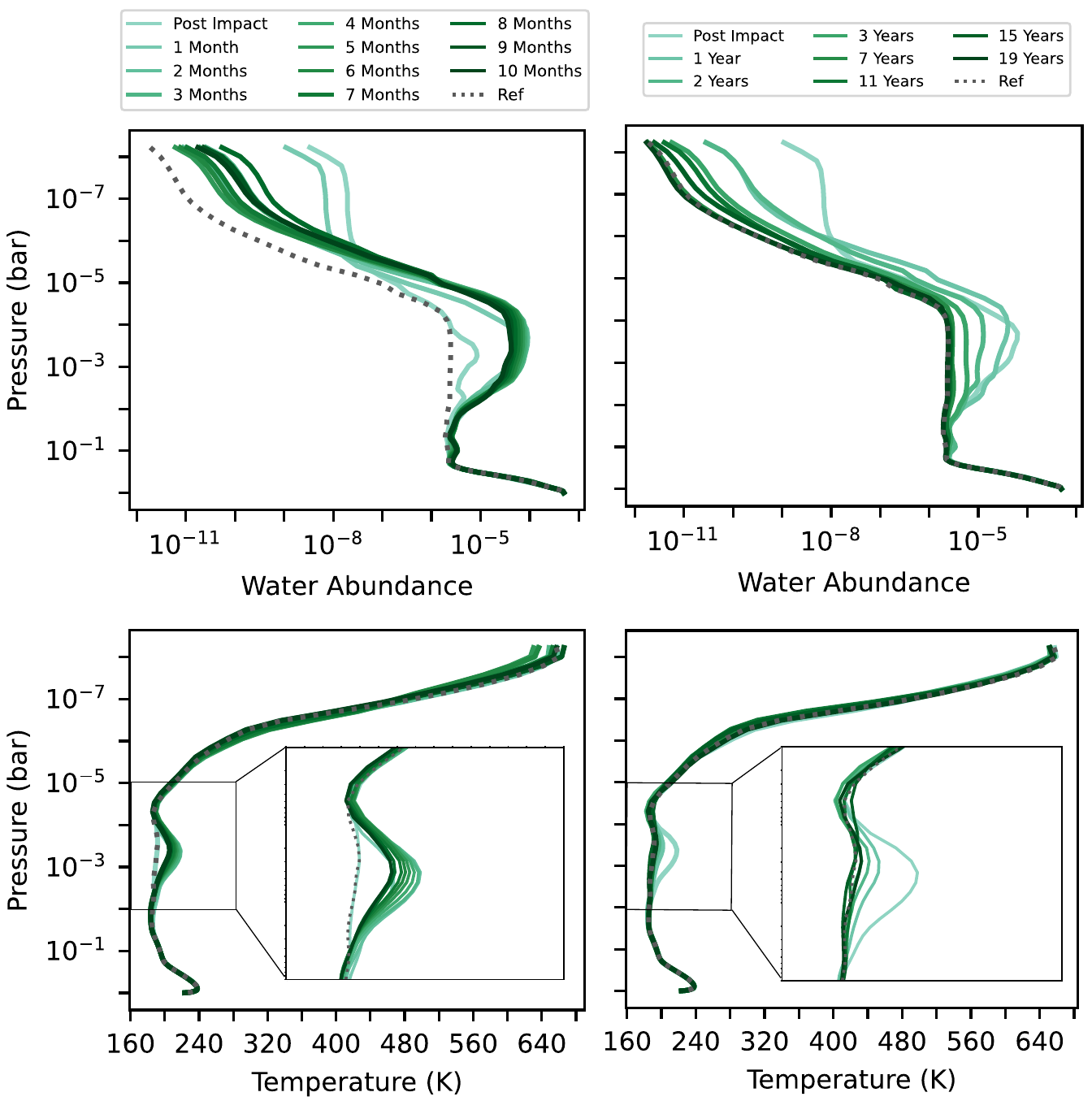}
\caption{Fractional water abundance (top) and temperature (bottom) profiles showing the rapid atmospheric evolution within the first 10 months (left) of the isolated deposition of water from an icy cometary impact, and the slower but steady settling of the atmosphere into a quasi-steady-state (right) reminiscent of the non-impacted reference state (grey dashed). Here each profile is calculated by averaging both horizontally over all latitudes and longitudes, and temporally over a month of simulation time. To better demonstrate the change in temperature in the mid-atmosphere, we include an inset showing a zoomed-in view of the temperature profile between $10^{-2}$ and $10^{-5}$ bar.   \label{fig:water_only_curves_water_temp} }
\end{centering}
\end{figure*}
\begin{figure*}[tbp]
\begin{centering}
\includegraphics[width=0.99\textwidth]{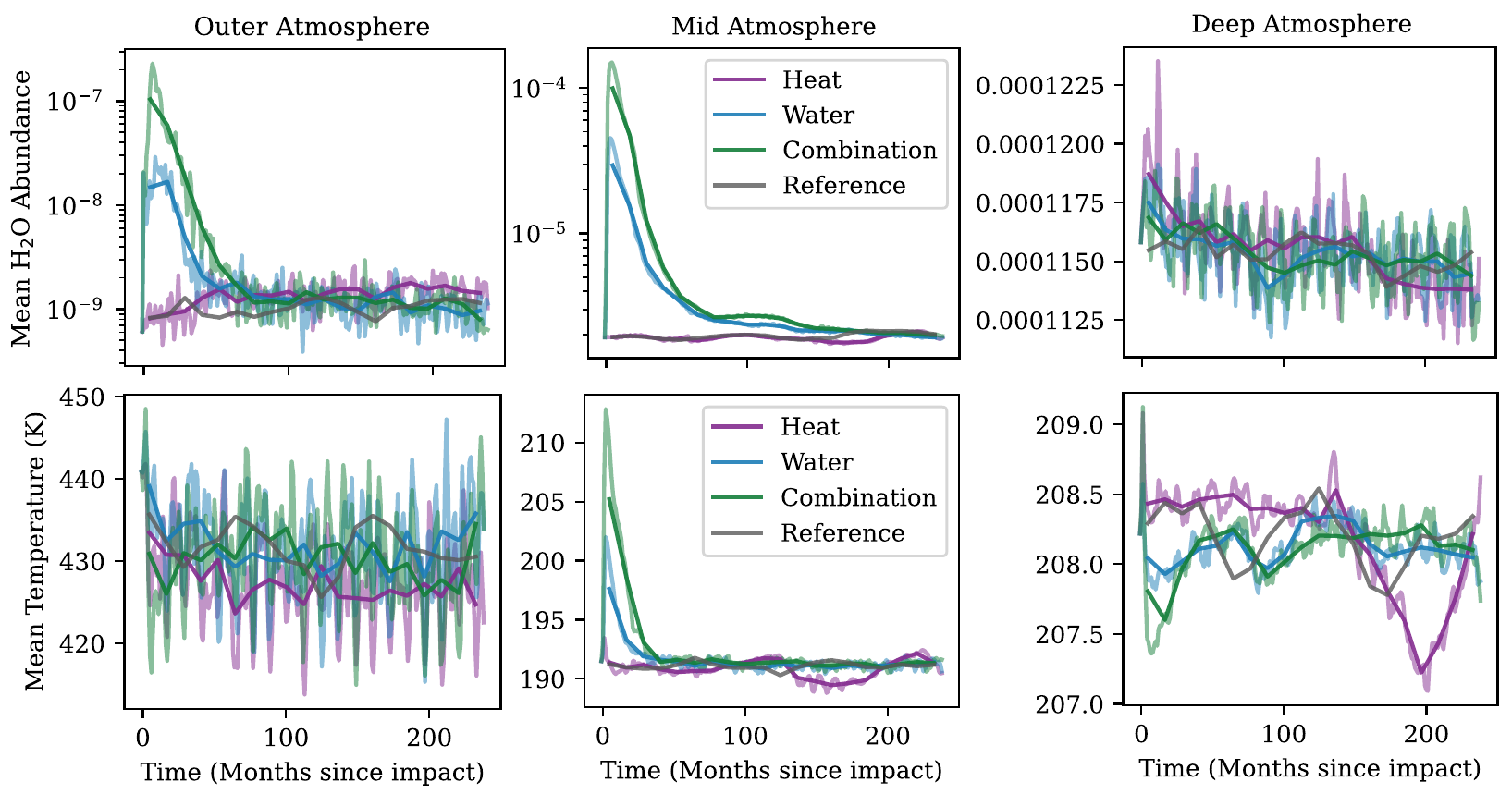}
\caption{ Time evolution of the annual mean (solid lines) and monthly mean (faint lines) fractional water abundance (top row) and temperature (bottom) in the outer atmosphere ($P<10^{-5}$ bar - left), mid-atmosphere ($10^{-5}>P>10^{-2}$ bar - middle), and near the surface ($P>10^{-2}$ bar - right) for all four models considered here: our water deposition only model (blue), thermal energy deposition only model (purple), {fiducial} model (i.e. combined water and thermal deposition - green), and non-impacted reference state (grey).    \label{fig:comparison_time_evo_curves_water_temp} }
\end{centering}
\end{figure*}

\appendix
\section{Results: Isolated Deposition} \label{sec:results_isolated}

To better {isolate how different components of the icy cometary impact influence the atmospheres of our tidally-locked, terrestrial, exoplanetary atmosphere, we also ran a pair of models in which we isolated/disentangled the effects of mass/water (red) and thermal energy (green/blue) deposition (\autoref{fig:energy_mass_deposition}). As was the case in our fiducial model, we consider the impact of} a pure water ice comet with a radius of $2.5$ km and a density of 1 gcm$^{-3}$ (\autoref{tab:comet_parameters}), and evolve the model for twenty years, within which time both models approach a quasi-steady-state in which the scale of the oscillations in the global mean temperature and fractional water abundance are similar to that found in our reference, non-impacted, atmosphere. 
By isolating the deposition in this way we can study the timescale and strength of the atmosphere's response to both components of the cometary material delivery independently, using this knowledge to better interpret our {fiducial model which introduced} both deposition profiles to our atmospheric model simultaneously (\autoref{sec:results_combined}).  
\subsection{Isolated Water Deposition} \label{sec:water_isolated} 
{We start by explore the isolated effects of cometary water delivery on our tidally-locked, Earth-like, atmosphere. For this isolated water deposition scenario, \autoref{fig:water_only_curves_water_temp} shows how the fractional water abundance (top row) and mean temperature (bottom) vary both shortly after impact (left column) and over the $\sim20$ years required for the model atmosphere to reach a quasi-steady-state (right column). As in our fiducial case, \autoref{sec:fiducial_evo}, we find that whilst most of the water is delivered to pressures $>10^{-4}$ bar, vertical transport carries this water aloft, where it persists as a multi-order-of-magnitude enhancement for at least five years post impact.
However, as shown in \autoref{fig:comparison_time_evo_curves_water_temp}, the enhancement in both the mid-atmosphere (top-middle) and outer-atmosphere (top-left) water abundance (blue) is weaker and { very} slightly shorter-lived than in our fiducial model (green), the latter of which combines the effects of both water/mass and thermal energy deposition. As we discuss below (\autoref{sec:heat_isolated}), the underlying cause of this is the thermal energy from the cometary impact driving water/ice evaporation/sublimation, { particularly in the mid-atmosphere}. \\
In turn, as was found for our fiducial model, this impact enhanced mid-atmosphere water vapour acts as a strong source of opacity, absorbing incoming radiation and driving local heating. However the lower peak fractional water abundance when compared with our fiducial model in turn drives a weaker local heating effect: at its peak in the resolved, i.e. non-time-averaged, data we find a mid-atmosphere temperature inversion which is { $\sim25$} K hotter than the same location in our non-impacted reference case (shown in grey). This is {$\sim10$} K cooler than the same peak found in our fiducial, combined deposition case. A similar story holds true for the average mid-atmosphere temperature, which is { $\sim 10$} K hotter than our non-impacted reference state whilst also, again, being { $\sim10$} K cooler than our fiducial model. \\

Differences between the isolated water deposition and fiducial models are also apparent in the deep atmosphere. As seen in the top-right panel of \autoref{fig:comparison_time_evo_curves_water_temp}, { our isolated water deposition profile exhibits a slightly stronger post-impact water enhancement than our fiducial case. }The underlying driver of this is a slight cooling of the deep atmosphere (bottom-right panel of \autoref{fig:comparison_time_evo_curves_water_temp}), driven by the absorption of incoming irradiation in the mid-atmosphere { (which is stronger in our fiducial case due to the enhanced water vapour content)}, which leads to an enhanced formation of snow/ice. \\

Overall, we find that whilst the changes driven by the isolated delivery of cometary water are similar to those seen in our fiducial model, differences are also abundantly clear. To better understand these differences, we also explored the isolated effects of impact-driven thermal energy deposition on our tidally-locked Earth-like atmosphere.

\begin{figure*}[tbp]
\begin{centering}
\includegraphics[width=0.9\textwidth]{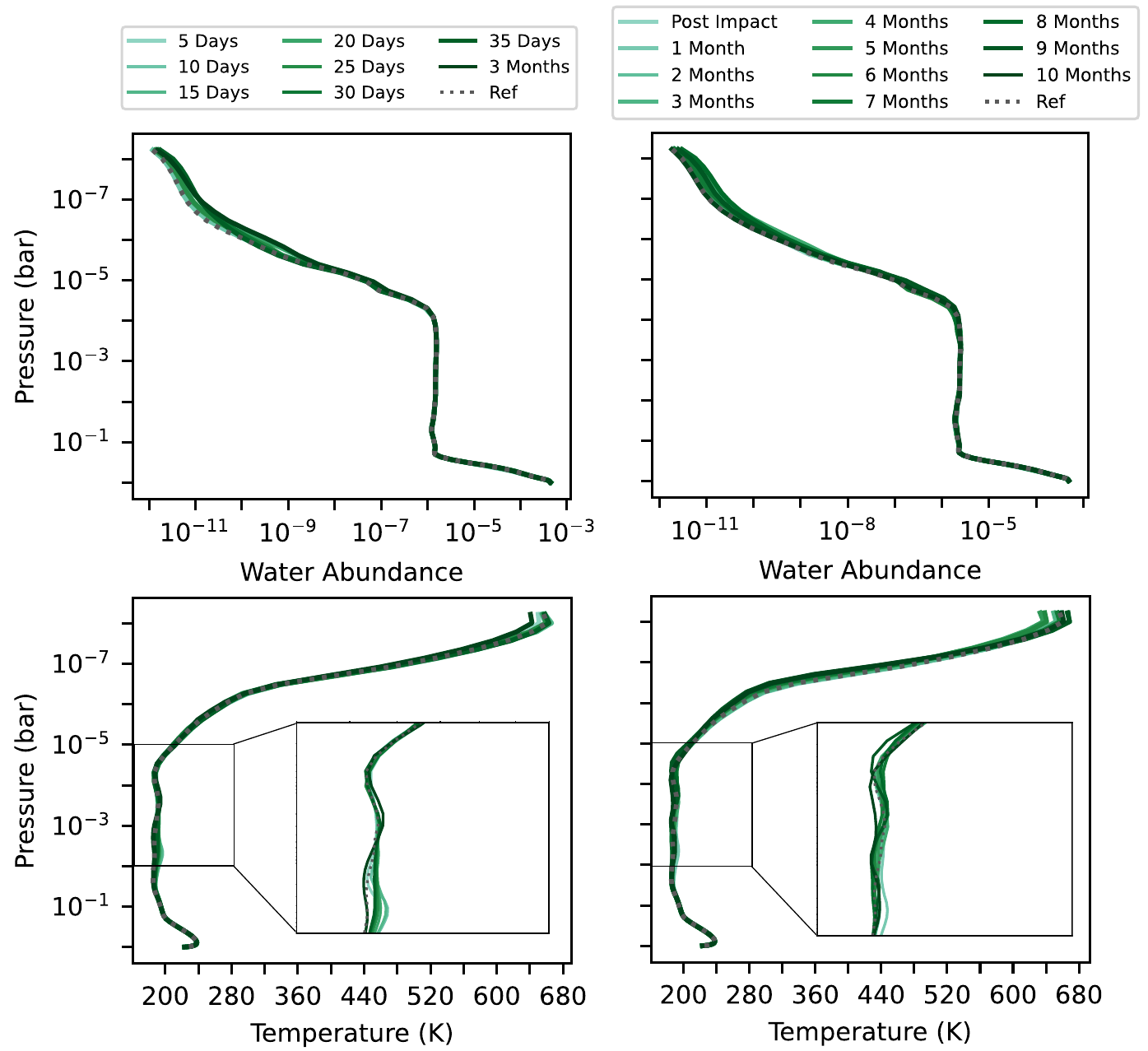}
\caption{ Fractional water abundance (top) and temperature (bottom) profiles showing the limited long-term effects of the isolated heat deposition from the impact of a pure water ice comet. Here we plot the horizontally averaged, over all latitudes and longitudes, profiles within the first 3 months (taken from 5 day means - left) and first 10 months (taken from monthly means - right) of the impact, and compare them to our non-impacted reference state (grey dashed). To better demonstrate the change in temperature in the mid-atmosphere, we include an inset showing a zoomed-in view of the temperature profile between $10^{-2}$ and $10^{-5}$ bar. Note the difference in timescale when compared with our water deposition and combined deposition simulations. \label{fig:heat_only_curves_water_temp} }
\end{centering}
\end{figure*}
\subsection{Isolated Thermal Energy Deposition} \label{sec:heat_isolated}
    
{As might be expected due to differences in typical radiative (weeks to months) and dynamical (i.e. chemical mixing/transport - months to years) timescales, the atmospheric response to the thermal energy deposition is much shorter lived than the changes associated with cometary water. This can be seen in both \autoref{fig:heat_only_curves_water_temp}, which shows how the fractional water abundance (top row) and mean temperature (bottom) vary both in the very first days post impact (left) and over the $<1$ year required for radiative effects to fully dissipate (right column), and the purple time evolution curves of \autoref{fig:comparison_time_evo_curves_water_temp}. \\
For example, the { slight} temperature enhancement in the mid-atmosphere, where most of the kinetic energy is deposited due to the cometary break-up, almost completely dissipates within a month of the impact. The remainder, $<1$ K with respect to our non-impacted reference state, dissipates over the following months.\\ 
A similar story holds true in both the outer and deep atmosphere, however in both of these regions we also find that the heating has a knock on effect on the atmospheric composition: a { slight} increase in the fractional water abundance relative to our non-impacted reference state. This occurs because the thermal energy from the cometary impact causes some of the frozen (ice/snow) and liquid (rain) water to sublimate/evaporate, leading to a weak local enhancement in the fractional water abundance, particularly near the surface (see the top-right panel of \autoref{fig:comparison_time_evo_curves_water_temp}). Whilst these changes to the water vapour content of the atmosphere are small relative to those associated with the cometary mass (water) deposition, our fiducial model reveals how the inclusion of thermal energy deposition in a combined deposition model results in a distinct climate from that found when considering water deposition alone. }

\begin{acknowledgements}
\nolinenumbers
The authors would like to thanks Richard Anslow for pointing out the error in \autoref{eq:ablation}. \\
F. Sainsbury-Martinez and C. Walsh would like to thank UK Research and Innovation for support under grant number MR/T040726/1. Additionally, C. Walsh would like to thank the University of Leeds and the Science and Technology Facilities Council for financial support (ST/X001016/1). This work was undertaken on ARC4, part of the High Performance Computing facilities at the University of Leeds, UK.\\
\end{acknowledgements}

\bibliography{papers}{}
\bibliographystyle{aasjournal}

\end{document}